\documentclass[aps,twocolumn,prb,preprintnumbers,amsmath,amssymb]{revtex4}
\usepackage{graphicx, bm}
\usepackage{dcolumn}
\usepackage{float}
\usepackage{latexsym}
\usepackage{amsmath}
\usepackage{graphics}
\usepackage{amssymb}
\usepackage{layout}
\usepackage{verbatim}
\usepackage{amsfonts,epsfig}
\usepackage{color}
\usepackage{setspace}
\newcommand{\beqnar}{\begin{eqnarray}}
\newcommand{\eeqnar}{\end{eqnarray}}

\newcommand{\bk}{{\bf k }}

\newcommand{\bq}{{\bf q }}
\newcommand{\br}{{\bf r }}

\newcommand{\beq}{\begin{equation}}
\newcommand{\eeq}{\end{equation}}
\newcommand{\enq}{\end{equation}}
\newcommand{\nav}    {\langle n\rangle}
\newcommand{\nrms}   {n_{\rm rms}}
\newcommand{\ceq}[1] {(\ref{#1})}
\newcommand{\rb}     {r_{\rm sc}}

\begin{document}
\title{Effect of charged impurity correlation on transport in monolayer and bilayer graphene}
\author{Qiuzi Li$^1$, E. H.\ Hwang$^1$, E. Rossi$^2$}
\affiliation{$^1$Condensed Matter Theory Center, Department of Physics, University of Maryland, College Park, Maryland 20742\\
             $^2$Department of Physics, College of William and Mary, Williamsburg, VA 23187, USA}
\date{\today}
\begin{abstract}
We study both monolayer and bilayer graphene transport properties taking into
account the presence of correlations in the spatial distribution of
charged impurities.
In particular we find that the experimentally observed
sublinear scaling of the graphene conductivity can be naturally explained
as arising from impurity correlation effects in the Coulomb
disorder, with no need to assume the presence of short-range scattering
centers in addition to charged impurities.
We find that also in bilayer graphene correlations among impurities
induce a crossover of the scaling of the conductivity at higher carrier densities.
We show that in the presence of correlation among charged impurities
the conductivity depends nonlinearly on the impurity density $n_i$ and
can even increase with $n_i$.
\end{abstract}


\maketitle

\section{Introduction}
The scaling of the  conductivity $\sigma$ as a function of
gate-voltage, proportional to the average carrier density $n$,
is invaluable in characterizing the properties of graphene
\cite{novoselov2004}.
The functional dependence of $\sigma(n)$  at low temperatures
contains information \cite{dassarma2010,peres_RMP10} about the nature of disorder
in the graphene environment (i.e., quenched charged impurity
centers, lattice defects\cite{FuhrerDefect_PRL}, interface roughness
\cite{Ishigami_Nano07},
ripples\cite{katsnelson2008,Bao_NatTech09}, resonant
scattering centers
\cite{Stauber_resonantPRB07,Monteverde_PRL10,wehiling2010c,Ferreira_PRB11},
etc.)
giving rise to the dominant scattering mechanism.
At finite temperatures electron-phonon scattering contributes to the resistivity
\cite{EfetovKim_PRL10,HwangDasPhonon_PRB08,MinHwangDas_arX10}. However,
in graphene the electron-phonon scattering is very weak and it becomes important
only at relatively high temperatures ($\gtrsim 400 K$), as evidence also from the fact that
around room temperature
the temperature dependence of $\sigma$ appears to be dominated
by activation processes \cite{qzli_PRB11,Heo_PRB11}.
The quantitative
weakness of the electron-phonon interaction in graphene gives
particular impetus to a thorough understanding of the disorder
mechanisms limiting graphene conductivity since this may enable
substantial enhancement of room temperature graphene-based device
for technological applications.
This is in sharp contrast to other high-mobility 2D
systems such as GaAs-based devices whose room-temperature mobility
could be orders of magnitude lower than the corresponding
low-temperature disorder-limited mobility due to strong carrier
scattering by phonons\cite{HwangDasGaAsmobi_PRB08}.
Therefore, a complete understanding of
the disorder mechanisms controlling $\sigma(n)$ in graphene at $T=0$
is of utmost importance both from a fundamental and a technological
prospective.

The experimental study of $\sigma(n)$ in gated graphene goes back to
the original discovery of 2D graphene,\cite{novoselov2004,Novoselov2}  and is a
true landmark in the physics of electronic materials.
Essentially, all experimental work on graphene
begins with a characterization of $\sigma(n)$ and the mobility,
$\mu=\sigma/(ne)$. A great deal is
therefore known
\cite{novoselov2004,Novoselov2,TanDas_PRL07,ChenJ_NPH_2007,BolotinYacoby,Feldman2} about the
experimental properties of $\sigma(n)$ in graphene.
The most important
features of the experimentally observed $\sigma(n)$ \cite{Novoselov,Novoselov2,TanDas_PRL07,ChenJ_NPH_2007,BolotinYacoby,Feldman2,ZhuExp_PRB09}
in monolayer graphene (MLG) are: (1) a non-universal sample-dependent minimum conductivity
$\sigma(n\approx0)\equiv\sigma_{min}$ at the charge neutrality point
(CNP) where the average carrier density vanishes;
(2) a linearly increasing,
$\sigma(n)\propto n$ , conductivity with increasing carrier density on
both sides of the CNP upto some sample dependent characteristic
carrier density; (3) a sublinear $\sigma(n)$ for high carrier density,
making it appear that the very high density $\sigma(n)$ may be
saturating.

To explain the above features of $\sigma(n)$ a model has been proposed
\cite{dassarma2010,Adam,Rossi,HwangAdamDas_PRL07,AndoMac,NomuraPRL}
with two distinct scattering mechanisms:
the long-range Coulomb
disorder due to random background charged impurities and static
zero-range (often called ``short-range") disorder.
The net graphene conductivity with these two scattering
sources
is then given by $\sigma\equiv\rho^{-1}=(\rho_c +
\rho_s)^{-1}$, where $\rho_c$
and $\rho_s$ are resistivities arising respectively from charged
impurity and short-range disorder.
It has been shown that
\cite{dassarma2010,Adam,Rossi,HwangAdamDas_PRL07,AndoMac,NomuraPRL}
$\rho_c\sim1/n$ and $\rho_s\sim$ constant in graphene, leading to
$\sigma(n)$ going as
\begin{equation}
\sigma(n) = \frac{n}{A + C n},
\label{eq:sig1}
\end{equation}
where the density independent constants $A$ and $C$ are known \cite{dassarma2010} as
functions of disorder parameters; $A$, arising from Coulomb disorder,
depends on the impurity density ($n_i$)
(and also weakly on their locations in space)
and the background dielectric constant ($\kappa$)
whereas the constant $C$, arising from the short-range disorder
\cite{dassarma2010,HwangAdamDas_PRL07}, depends on the strength of the white-noise disorder characterizing the zero-range scattering.
Eq.~(\ref{eq:sig1}) clearly manifests the observed $\sigma(n)$ behavior of graphene for $n\neq 0$ since $\sigma(n\ll A/C)\sim n$, and $\sigma(n\gg A/C)\sim1/C$ with $\sigma(n)$ showing sublinear
$(C+A/n)^{-1}$ behavior for $n\sim A/C$.

The above-discussed scenario for disorder-limited graphene
conductivity, with both long-range and short-range disorder playing
important qualitative roles at intermediate $(n_i\lesssim n\leqslant
A/C)$ and high $(n>A/C)$ carrier densities respectively, has been
experimentally verified by several groups
\cite{TanDas_PRL07,ChenJ_NPH_2007,BolotinYacoby,Feldman2,ZhuExp_PRB09}.
There is, however, one {\it serious issue } with this reasonable scenario: although the physical mechanism underlying the long-range
disorder scattering is experimentally established \cite{dassarma2010,TanDas_PRL07,ChenJ_NPH_2007} to be the presence of
unintentional charged impurity centers in the graphene environment,
the physical origin of the short-range disorder scattering is unclear
and has so far eluded direct imaging experiments. As a matter of fact the experimental evidence suggests that
point defects (e.g. vacancies) are rare
in graphene and should produce negligible short-range disorder. There have also been occasional puzzling conductivity measurements
[e.g., Ref.~\onlinecite{GeimIce_PRL09,Schedin_NM09}] reported in the literature which do not appear to be explained by
the standard model of independent dual scattering by long- and short-range disorder playing equivalent roles.

Recently a novel theoretical model has been
proposed \cite{qzli_PRL11} that is able to semiquantitatively explain all the major
features of $\sigma(n)$ observed experimentally assuming only the presence
of charged impurities. The key insight on which the model relies is the fact that
in experiments, in which the samples are prepared at room temperature and are often also current annealed,
it is very likely that spatial correlations are present among
the charged impurities. In particular this model is able to explain
the linear (sublinear) scaling of $\sigma(n)$ in MLG at low (high) $n$ without
assuming the presence of short-range scattering centers.

In this work we first review the transport model proposed in Ref. [\onlinecite{qzli_PRL11}],
and then extend it to the case of bilayer graphene (BLG).
We find that, as in MLG, the presence of spatial-correlations among impurities
is able to explain a crossover of the scaling of $\sigma(n)$ from low $n$ to high $n$
in BLG, as observed in experiments, and that, because of the spatial
correlations, $\sigma$ depends non-monotonically on the impurity density $n_i$.

The remainder of this paper is structured as follows. In
Section \ref{sec:strufac} we present the model and the results for
the structure factor $S(\bq)$ that
characterizes the impurity correlations.
With the structure factor calculated
in Sec.~\ref{sec:strufac} we provide the
transport theory  in Section \ref{sec:mlg} and Section
\ref{sec:blg}. In Section \ref{sec:mlg}, we study the density-dependent conductivity $\sigma (n)$ of monolayer graphene in the
presence of correlated charged impurities. We calculate $\sigma(n)$ at
higher carrier density using the Boltzmann transport theory. We also
evaluate $\sigma(n)$ applying both Thomas-Fermi-Dirac theory \cite{rossi2008} and
effective medium theory \cite{Rossi} to characterize the strong carrier density
inhomogeneities close to the charge neutrality point. In Section
\ref{sec:blg}, we apply the Boltzmann transport theory and the
effective medium theory for
correlated disorder to bilayer graphene and discuss the
qualitative similarities and the quantitative differences between
monolayer and bilayer graphene. We briefly review the experimental
situation in Section \ref{sec:discuss}. We then conclude in Section
\ref{sec:conclu}.

%

%
%

%
%
%
%
%

%

%

%
%
%

%
%


\section{Structure factor $S(\bq)$ of Correlated disorder}
\label{sec:strufac}
In this section we describe the model used to calculate the
structure factor $S(\bq)$ for the charged impurities.
We then present results for $S(\bq)$ obtained using this model via
Monte Carlo simulations. The Monte Carlo results are then used to
build a simple continuum approximation for $S(\bq)$, which
captures all the features of $S(\bq)$ that are relevant for
the calculation of $\sigma(n)$.

\subsection{Model for the structure factor $S(\bq)$}

To calculate $S(\bq)$ we follow the procedure presented
in Ref. \onlinecite{Kawamura_SCC96}, adapted to the case
of a honeycomb structure.
The approach  was
applied to study the effects of impurity scattering in GaAs
heterojunctions and successfully explained the experimental observation of
high-mobilities (e.g. greater than $10^7$ cm$^2$/(V$\cdot$s)) in
modulation-doped GaAs heterostructures.
The possible charged impurity positions on graphene form a triangular lattice specified
by $\mathbf{r}_{LM} = \mathbf{a} L + \mathbf{b} M $. The vectors $\mathbf{a}=(1,0) a_0$
and  $\mathbf{b}=(\sqrt{3}/2,1/2) a_0$ defined in the x-y plane, with $a_0 =4.92${\AA},
which is two times the graphene lattice constant since the most densely packed phase
of impurity atoms (e.g. K as in Ref. \onlinecite{ChenJ_NPH_2007}) on graphene is likely to be an
$m\times m$ phase with $m=2$ for K \cite{Caragiu_JPCM05}.
The structure factor, including the Bragg scattering term, is given by the following equation:
\begin{equation}
 S(\bq) =\frac{1}{N_i}\langle \sum_{i,j}  e^{i \bq\cdot(\br_i-\br_j} \rangle
 \label{eq:sq1}
\end{equation}
where $\br_i, \br_j$ are the random positions on the lattice $\bf r_{LM}$ of the charged impurities
and the angle brackets denote averages over disorder realizations.
Introducing the fractional occupation $f\equiv N_i/N$ of the total
number of available lattice sites $N$ by the number  of charged
impurities $N_i$, and the site occupation factor $\epsilon_{LM}$ equal to
1 if site $\br_l$ is occupied or zero if unoccupied,
we can rewrite Eq. (\ref{eq:sq1}) as
\begin{equation}
 S(\bq) = \frac{1}{f}\sum_{LM} \langle \epsilon_{LM}\epsilon_0\rangle e^{i\bq\cdot\br_{LM}}
 \label{eq:sq2}
\end{equation}
in which the sum is now over all the available lattice sites (not only the ones
occupied by the impurities).
By letting $C_{LM} \equiv \langle \epsilon_{LM}\epsilon_0\rangle/f^2$
we can rewrite Eq. (\ref{eq:sq2}) as:
\begin{equation}
 S(\bq) = f \sum_{LM} C_{LM} e^{i\bq\cdot\br_{LM}}.
 \label{eq:sq3}
\end{equation}
We then subtract  the Bragg scattering term from this expression
considering that it does not contribute to the resistivity obtaining
\begin{equation}
 S(\bq) = f \sum_{LM} (C_{LM} - 1) e^{i\bq\cdot\br_{LM}}.
 \label{eq:sq4}
\end{equation}
It is straightforward to see that for the totally random case, the structure factor is given by $S(\mathbf{q})=1-f$
and $n_{i} \simeq 4.8f\times10^{14}\text{cm}^{-2}$.
For the correlated case we assume that two impurities cannot be closer than a
given length $r_0<r_i\equiv (\pi n_i)^{-1/2}$ defined as the correlation length. This model is motivated by the fact that
two charged impurities cannot be arbitrarily close to each
other because the Coulomb repulsion among the impurities during device growth
and there must be a minimum separation between them.

\subsection{Monte Carlo results for $S(\bq)$}
Using Monte Carlo simulations carried out on a $200\times200$ triangular lattice with $10^6$ averaging runs and periodic boundary conditions
we have calculated the structure factor given by Eq. (\ref{eq:sq4}).
In the Monte Carlo calculation a lattice site is
chosen randomly and becomes occupied only if it is initially unoccupied and has no nearest neighbors within the correlation length $r_0$.
This process is repeated until the required fractional occupation for a given impurity density is obtained. Once the configuration is generated, the $C_{LM}$ can be numerically determined after doing the ensemble average. In the numerical calculations, we use only statistically
significant $C_{LM}$, i.e., $|\mathbf{r}_{LM}-\mathbf{r}_{00}|\leq3 r_0$, since $C_{LM}$ is essential unity for $|\mathbf{r}_{LM}-\mathbf{r}_{00}| > 3 r_0$.

\begin{figure}
\includegraphics[width=0.99\columnwidth]{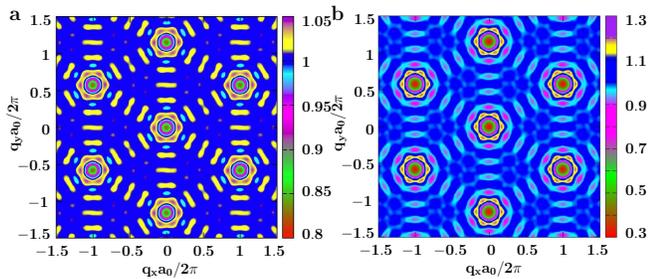}
\caption{
 (a) Density plot of the structure factor $S(\bq)$ obtained from Monte
  Carlo simulations
  for $a_0=4.92\;\AA$ and $r_0=5a_0$. (a) $n_i = 0.95\times10^{12}$~cm$^{-2}$; (b) $n_i = 4.8\times10^{12}$~cm$^{-2}$.
}
\label{fig:DensPlotStrufac}
\end{figure}

In Fig.~\ref{fig:DensPlotStrufac}, we present a contour plot of the
structure factor $S(\bq)$ obtained from the
Monte Carlo simulations for two different values of the impurity density.
For $r_0\neq 0$ the structure factor is suppressed at small momenta.
Moreover the suppression of $S(\bq)$ at small momenta is more pronounced, for
fixed $r_0$, as $n_i$ is increased as it can be seen comparing
the two panels of Fig.~\ref{fig:DensPlotStrufac}.
The magnitude of $S(\bq)$ at small $\bq$ mostly determines the d.c.
conductivity and therefore, from the results of Fig.~\ref{fig:DensPlotStrufac},
is evident that the presence of spatial correlations among the charged impurities
will strongly affect the value of the conductivity.


\subsection{Continuum model for $S(\bq)$}

\begin{figure}
\includegraphics[width=0.99\columnwidth]{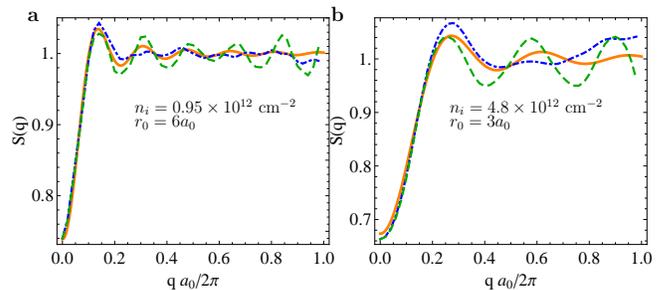}
\caption{ (a) and  (b) show the calculated structure factor $S(\bq)$ for two values of impurity density $n_i$. (a) $n_i = 0.95 \times 10^{12}$ cm$^{-2}$; (b) $n_i = 4.8 \times 10^{12}$ cm$^{-2}$. The solid lines show $S(\bq)$  using Eq.~(\ref{eq:strufac2}).  Dot-dashed and dashed lines show the Monte Carlo results for two different directions of $\bq$ from $x$-axis, $\theta = 0$ and $\theta=30^{\circ}$, respectively.
}
\label{fig:2Dstrufac}
\end{figure}

Given that the value of the d.c. conductivity depends almost entirely on the
value of $S(\bq)$ at small momenta, as discussed in Sections
\ref{sec:mlg} and \ref{sec:blg}, it is convenient to introduce a simple
continuum model being able to reproduce for small $\bq$ the structure factor
obtained via Monte Carlo simulations.
A reasonable continuum approximation to the above discrete lattice model
is given by the following pair distribution function $g({\br})$
(${\br}$ is a 2D vector in the graphene plane),
\begin{align}
g(\br) =\begin{cases}
0 & |\br| \leq r_0\\
1 & |\br| > r_0 \end{cases}.
\label{eq:g}\end{align}
for the impurity density distribution.
In terms of the pair correlation function $g(\br)$ the structure factor is given by:
\begin{equation}
S(\bq)  =1+n_{i}\int d^{2}r e^{i\mathbf{q\cdot r}}[g(\mathbf{r})-1]
\label{eq:strufac1}
\end{equation}
For uncorrelated random impurity scattering, as in the standard theory,
$g(\br)=1$ always, and $S(\bq)\equiv 1$.
With Eqs.~(\ref{eq:g}) and ~(\ref{eq:strufac1}), we have
\begin{equation}
S(q) =1-2\pi n_{i}\frac{r_{0}}{q}J_{1}(qr_{0})
\label{eq:strufac2}
\end{equation}
where $J_1(x)$ is the Bessel function of the first kind.
Fig. \ref{fig:2Dstrufac} shows $S(\bq)$
obtained both via Monte Carlo simulations and by using
the simple continuum analytic model [Eq.~(\ref{eq:strufac2})]
for a few values of $r_0$ and $n_{i}$.
We can see that the continuum model reproduces extremely well
the dependence of the structure factor on $\bq$ for small
momenta, i.e. the region in momentum space that is relevant
for the calculation of $\sigma$.


\section{Monolayer graphene conductivity}
\label{sec:mlg}
In this section, we explore how the spatial correlations among charged impurities affect monolayer graphene transport properties.
To minimize the parameters entering the model we assume the charged impurities to be in a 2D plane
placed at an effective distance $d$ from the graphene sheet (and parallel to it).

We first study the density-dependent conductivity in monolayer graphene transport for large carrier densities ($n \gg n_i$) using the Boltzmann transport theory, where the density fluctuations of the system can be ignored. We then discuss
$\sigma(n)$ close to the charge neutrality point, where the graphene landscape breaks up
into puddles \cite{martin2008, rossi2008, YZhang_NP09,LeRoyPuddle_PRB09,Adam_SCC09,LeRoyPuddle_PRB11} of electrons and holes due to the effect of the charged impurities using the effective medium theory developed in Ref.[\onlinecite{Rossi}].

\subsection{High density: Boltzmann transport theory}
\label{subsec:mlgbtt}

Using the Boltzmann theory for the  carrier conductivity at temperature $T=0$ we have
\beq
 \sigma=\frac{e^2}{h}\frac{gE_F\tau(E_F)}{2\hbar},
 \label{eq:sigma_boltz}
\eeq
where $E_F$ is the Fermi
energy, $g=4$ is the total degeneracy of
graphene, and $\tau$ is the transport relaxation time at the Fermi
energy obtained using the Born approximation.
The scattering time at $T=0$ due to the disorder potential created by charged impurities taking into account the spatial correlations among impurities is given by
\cite{HwangDas_PRB07,qzli_PRB11,note01}:
\begin{eqnarray}
\dfrac{\hbar}{\tau(\epsilon_{p{\bf k}})} &=& 2\pi n_{i} \int \dfrac{d^2 k'}{(2\pi)^2}\left[\dfrac{V(|\mathbf{k-k'}|)}{\varepsilon(|\mathbf{k-k'}|)}\right]^{2} S(\mathbf{k-k'}) \nonumber \\
& \times & g(\theta_{\bf kk'}) \left[1-\cos\theta_{\bf kk'}\right]\delta(\epsilon_{p\mathbf{k'}}-\epsilon_{p\mathbf{k}})
\label{eq:mscatt}
\end{eqnarray}
where $V(q)=2\pi e^2/\kappa q e^{-q d}$ is the Fourier transformation of
the 2D Coulomb potential created by a single charged impurity
in an effective background dielectric
constant $\kappa$, $\varepsilon(q)$ is the static dielectric function,
$\epsilon_{s\mathbf{k}} = s \hbar v_F k$ is the carrier energy
for the pseudospin state ``$s$", $v_F$ is graphene Fermi velocity,
${\bf k}$ is the 2D wave vector, $\theta_{\bf kk'}$ is
the scattering angle between in- and out- wave vectors
${\bf k}$ and ${\bf k'}$, $g(\theta_{\bf kk'})=\left[1+\cos\theta_{\bf
    kk'}\right]/2$ is a wave function form-factor associated with the
chiral nature of MLG (and is determined by its band
structure).
The two dimensional static dielectric function
$\varepsilon(q)$ is calculated within the random phase
approximation (RPA) \cite{HwangDas_PRB07}, and given by
\begin{equation}
\varepsilon(q)=\begin{cases}
1+\dfrac{4k_{F}r_{s}}{q} & \text{if\ \ }q<2k_{F}\\
1+\dfrac{\pi r_{s}}{2} & \text{if}\ \ q>2k_{F}\end{cases}\label{eq:diele}
\end{equation}

After simplifying Eq. \ref{eq:mscatt},  the relaxation time in the presence of correlated disorder is given by:
\begin{equation}
\dfrac{\hbar}{\tau}=\Bigg(\dfrac{\pi n_{i}\text{\ensuremath{\hbar}}
v_{F}}{4k_{F}}\Bigg)r_{s}^{2}\int
\frac{d\theta \left(1-\cos^{2}\theta\right)}
{\left(\sin\frac{\theta}{2}+2r_{s}\right)^2}
S(2k_{F}\sin\frac{\theta}{2}),
\label{eq:relatime}
\end{equation}
where $k_F$ is the Fermi
wavevector ($k_F=E_F/(\hbar v_F)$), and $r_s$ is the graphene fine structure
constant ($r_s=e^2/(\hbar v_F\kappa) \simeq 0.8$ for graphene on a SiO$_2$ substrate).
For uncorrelated random impurity scattering (i.e.,
$r_0 = 0$, $g(\br)=1$, and $S(\bq)\equiv 1$) we recover the standard
formula for Boltzmann conductivity by screened random charged impurity
centers \cite{HwangAdamDas_PRL07,AndoMac,NomuraPRL}, where the
conductivity is a linear function of carrier density.

By approximating the structure factor $S(2k_F\sin\theta/2)$ that
appears in  (\ref{eq:relatime}) by a Taylor expansion
around $k_F\sin\theta/2=0$ it is possible to obtain an analytical expression
for $\sigma(n)$ that allows us to gain some insight on how the spatial
correlation among charged impurities affect the conductivity in MLG.
Expanding the first kind of Bessel function $J_1 (x)$ in
Eq. \ref{eq:strufac2} around $x \sim 0$ to the third order
\beq
J_{1}(x)\simeq\frac{x}{2}-\frac{x^{3}}{16}.
\eeq
from  Eq. (\ref{eq:relatime}) we obtain:
\begin{equation}
 \dfrac{\hbar}{\tau}
  \simeq \dfrac{4\pi n_{i}\text{\ensuremath{\hbar}}v_{F}}{k_{F}}r_{s}^{2}\left[G_{1}(r_s)\left(1-\pi n_{i}r_{0}^{2}\right)+G_{2}(r_s)\dfrac{\pi n_{i}k_{F}^{2}r_{0}^{4}}{2}\right],
 \label{eq:Anascatter}
 \end{equation}
where the dimensionless functions $G_{1}(x)$ and
$G_{2}(x)$ are given by, \cite{Hwang_PRB195412}
\begin{equation}
\begin{array}{l l l }
 G_{1}(x)  = \dfrac{\pi}{4}+6x-6\pi x^{2}+ 4x(6x^{2}-1)g(x),
 \\
 \\
G_{2}(x)  = \dfrac{\pi}{16}-\dfrac{4x}{3}+3\pi x^2
         +  40x^3  [1-\pi x+ \dfrac{4}{5} (5x^{2}-1)g(x) ],
\end{array}
\end{equation}
where
\begin{equation}
g(x)=\begin{cases}
\dfrac{\text{sech}^{-1}(2x)}{\sqrt{1-4x^2}} & \text{if\ \ }x<\frac{1}{2},
\\
\\
\dfrac{\text{sec}^{-1}(2x)}{{\sqrt{4x^{2}-1}}} & \text{if\ \ }x>\frac{1}{2}.
\end{cases}
\end{equation}
Using Eq.~(\ref{eq:sigma_boltz}), (\ref{eq:Anascatter}),
and recalling that $k_F=\sqrt{\pi n}$, we find:
\begin{equation}
\sigma(n)= \dfrac{A n}{ 1 - a + B a^2 n/n_i},
\label{eq:sigasym}
\end{equation}
where
\begin{eqnarray}
 A  &=&   \dfrac{e^{2}}{h}\frac{1}{2n_{i}r_{s}^{2}G_{1}(r_{s})} \nonumber \\
 a  &=&  \pi n_i r_0^2  \\
 B  &=&  \frac{G_2 (r_s)}{2G_1(r_s)}. \nonumber
\end{eqnarray}
Note $a<1$ in our model because the correlation length can not exceed
the average impurity distance, i.e., $r_0< r_i = (\pi
n_i)^{-1/2}$. Eq.~(\ref{eq:sigasym}) indicates that at low carrier
densities the conductivity increases linearly with $n$
at a rate that increases with $r_0$
\beq
\sigma(n)\sim \dfrac{A n}{(1-a)};
\eeq
whereas at large carrier densities the dependence of $\sigma$
on $n$ becomes sublinear:
\beq
\sigma(n)\sim 1-\dfrac{n_c}{n},
\eeq
where $n_c=(1-a)n_i/(Ba^2)\sim
O(1/n_ir_0^4)$. Note that the above equation is valid for $\sqrt{\pi
  n} r_0 \ll 1$, where we expand the structure factor as a power
series of $\sqrt{\pi n} r_0$. The crossover density $n_c$, where the
sublinearity
($n>n_c$) manifests itself, increases strongly with decreasing
$r_0$. This generally implies that the higher mobility annealed
samples should manifest stronger nonlinearity in $\sigma(n)$, since
annealing leads to stronger impurity correlations (and hence larger
$r_0$).
This behavior has been observed recently in experiments
in which the correlation among charged impurities
was controlled via thermal annealing \cite{fuhrer2010}.
Contrary to the standard-model  with no spatial correlation
among charged impurities in which the resistivity increases linearly in $n_i$,
Eq.~(\ref{eq:sigasym}) indicates that the resistivity could decrease
with increasing impurity density if there are sufficient inter-impurity
correlations.
This is due to the fact that, for fixed $r_0$,
higher density of impurities are more correlated
causing $S({\bf q})$ to be more strongly suppressed at low $q$ as
shown in Fig. \ref{fig:DensPlotStrufac} and \ref{fig:2Dstrufac}.
In the extreme case, i.e., $r_0=a_0$ and $r_i=r_0$,
the charged impurity distribution would be strongly correlated, indeed
perfectly periodic, and
the resistance, neglecting other scattering sources, would be zero.
From Eq.~(\ref{eq:sigasym}) we find that the resistivity reaches a maximum
when the condition
\begin{equation}
r_i/r_0 = \sqrt{2(1-\pi B n r_0^2)}.
\label{eq:optimal}
\end{equation}
is satisfied.
Equation~(\ref{eq:optimal}) can be used as a guide to improve the mobility of
graphene samples in which charged impurities are the dominant source
of disorder.

\begin{figure}
\includegraphics[width=0.99\columnwidth]{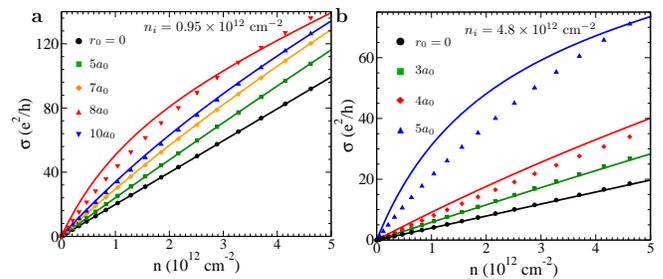}
\caption{
  Calculated $\sigma(n)$ in monolayer graphene with $S(\bq)$ obtained
  from the Monte Carlo
  simulations, symbols, and $S(\bq)$ given by Eq.~\ref{eq:strufac2}, solid lines for
  (a) $n_i=0.95\times 10^{12}$ cm$^{-2}$, and (b) $n_i=4.8\times 10^{12}$ cm$^{-2}$. The different lines correspond to different values of $r_0$, from top to bottom
  $r_0=10 a_0, 8 a_0, 7 a_0, 5 a_0, 0\;$ in (a) and $r_0=5 a_0, 4 a_0, 3 a_0, 0\;$ in (b).
}
\label{fig:MLGcond}
\end{figure}

Figs.~\ref{fig:MLGcond}(a) and (b) present the results for $\sigma(n)$
obtained integrating numerically the r.h.s. of Eq.~\ceq{eq:relatime}
and keeping the full momentum dependence of the structure factor.
The solid lines show the results obtained using the $S(\bq)$
given by the continuum model, Eq.~\ceq{eq:strufac2},
the symbols show the results obtained using the $S(\bq)$ obtained
via Monte Carlo simulations.
The comparison between the
two results shows that the analytic continuum correlation model is
qualitatively and quantitatively reliable.
It is clear that, {\it for the same value of $r_0$}, the
dirtier (cleaner) system shows stronger nonlinearity (linearity) in a
fixed density range consistent with the experimental observations
\cite{fuhrer2010} since the correlation effects are stronger for larger values of $n_i$.

\begin{figure}
\includegraphics[width=0.99\columnwidth]{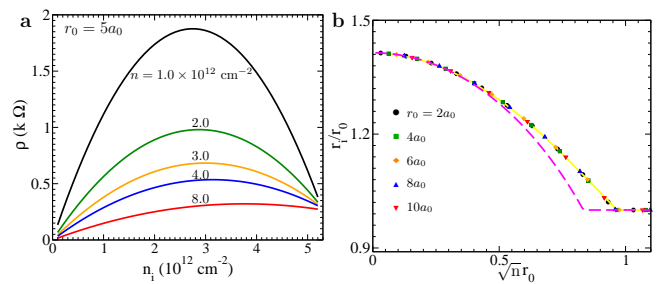}
\caption{(a) Calculated resistivity $\rho$ in monolayer graphene as a function of impurity density $n_i$ for different carrier densities
  with $r_0 =5 a_0$. (b) The relationship between $r_i/r_0$
  and $\sqrt{n} r_0$ in monolayer graphene, where the conductivity is minimum. The dashed line is obtained using Eq. \ref{eq:optimal}.
}
\label{fig:MLGoptimal}
\end{figure}

Fig. \ref{fig:MLGoptimal}(a) presents that the resistivity $\rho=1/\sigma$
in monolayer graphene as a function of impurity density
$n_i$ with correlation length $r_0 = 5 a_0$ for different values of
carrier density. It is clear that the impurity correlations  cause
a highly nonlinear resistivity as a function of impurity density
and that this nonlinearity in $\rho (n_i)$ is much stronger for lower
carrier density.
In Fig. \ref{fig:MLGoptimal}(b) we show the value of the ratio $r_i/r_0$
for which $\rho$ is maximum as a function of
$\sqrt{n} r_0$
The analytical expression of
Eq. \ref{eq:optimal} is in very good agreement with the result obtained
numerically using the full momentum dependence of $S(\bq)$.


\subsection{Low density: Effective medium theory}
\label{subsec:mlgemt}
\begin{figure}
\includegraphics[width=0.99\columnwidth]{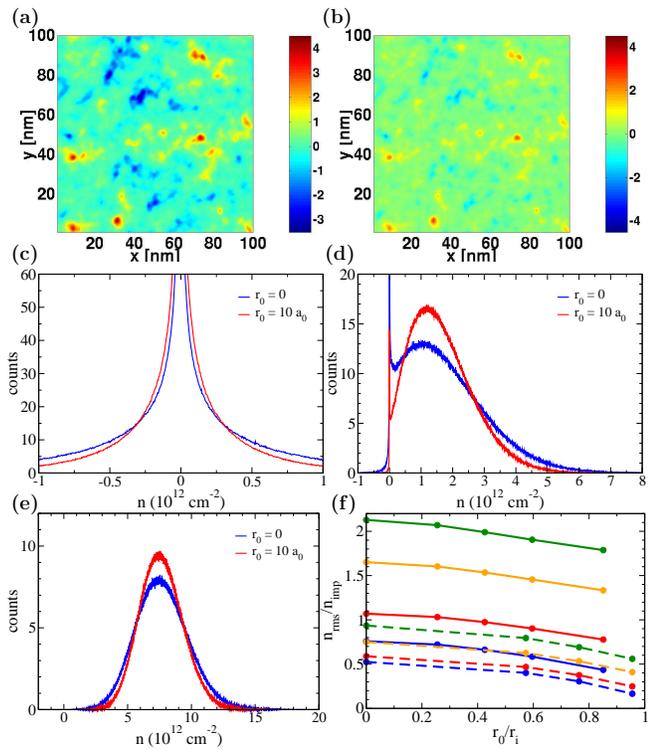}
\caption{(color online)
  The carrier density in monolayer graphene for a single disorder
  realization obtained from the TFD theory (a)
  for the uncorrelated case and (b) $r_0=10\;a_0$ with  $n_i
  =0.95\times10^{12}$~cm$^{-2}$.
  Carrier probability distribution function $P(n)$ are  shown in (c), (d), (e) for
  $\nav=0$, 1.78, $7.7\times 10^{12}$~cm$^{-2}$, respectively.
  In (f) the ratio $n_{\rm rms}/n_i$ is shown as a function of $r_0/r_i$
  for $n_i = 0.95\times10^{12}$ cm$^{-2}$, solid lines, and
  $n_i = 4.8\times10^{12}$ cm$^{-2}$, dashed lines.
We use $\nav=7.7$, 3.14, 0.94, $0\times 10^{12}$~cm$^{-2}$
  for the solid lines (from top to bottom) and
  $\nav=8.34$, 4.10, 1.7, $0\times 10^{12}$~cm$^{-2}$
  for the dashed lines.
}
\label{fig:2ER}
\end{figure}

Due to the gapless nature of the band structure,  the presence of
charged impurities induce strong carrier density inhomogeneities in MLG and BLG.
Around the Dirac point, the 2D graphene layer becomes a spatially inhomogeneous semi-metal with electron-hole puddles randomly located in the system. To characterize these inhomogeneities we use the Thomas-Fermi-Dirac (TFD) theory \cite{rossi2008}. Ref. [\onlinecite{Rossi}] has shown that the TFD theory coupled with the Boltzmann transport theory provides an excellent description of the minimum conductivity around the Dirac point with randomly distributed Coulomb impurities. We further improve this technique to calculate the density landscape and the minimum conductivity of monolayer graphene in the presence of correlated charged impurities. To model the disorder, we have assumed that the impurities are
placed in a 2D plane at a distance $d=1$~nm from the graphene layer. Fig.~\ref{fig:2ER}(a), (b) show the carrier density profile for a single disorder realization for the uncorrelated case and correlated case
($r_0=10\;a_0$) for $n_i =0.95\times10^{12}$~cm$^{-2}$. We can see
that in the correlated case the amplitude of the density fluctuations is much smaller than in
the uncorrelated case. The TFD approach is very efficient and allows the calculation of disorder averaged
quantities such as the density root mean square, $n_{\rm rms}$, and
the density probability
distribution $P(n)$. Figs. \ref{fig:2ER}(c), (d), (e) show $P(n)$ at
the CNP, and away
from the Dirac point ($n_i =0.95\times10^{12}$~cm$^{-2}$). In each
figure both the results for the uncorrelated case and the
one for the correlated case are shown. $P(n)$ for the correlated case is in general
narrower than $P(n)$ for the uncorrelated case resulting in smaller
values of $\nrms$ as shown in
Fig.~\ref{fig:2ER}(f) in which $\nrms/n_i$ as a function of $r_0/r_i$ is plotted
for different values of the average density, $\nav$, and two different
values of the impurity density, $n_i =0.95\times10^{12}$~cm$^{-2}$
(``low impurity density'') for the
solid lines, and $n_i =4.8\times10^{12}$~cm$^{-2}$ (``high impurity
density'') for the dashed lines.

\begin{figure}
\includegraphics[width=0.99\columnwidth]{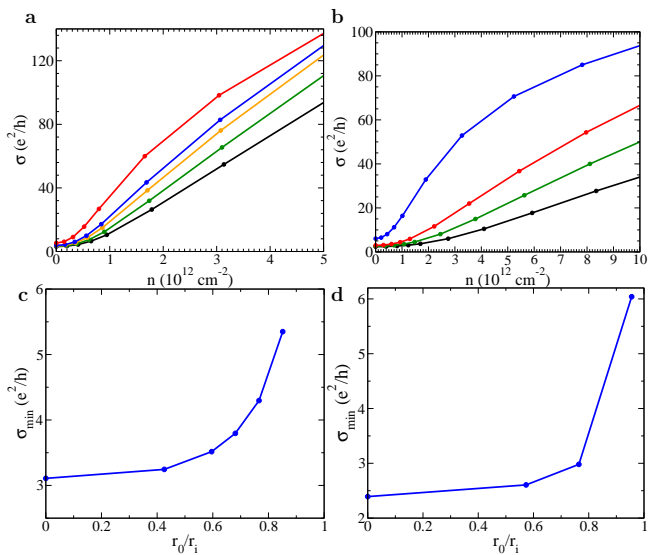}
\caption{
  (a) and (b) show the results for $\sigma(\nav)$ in monolayer graphene obtained from the EMT for $n_i=0.95\times 10^{12}$ cm$^{-2}$ and $n_i=4.8\times 10^{12}$ cm$^{-2}$ respectively. The different lines correspond to different values of $r_0$, from top to bottom
  $r_0=10 a_0, 8 a_0, 7 a_0, 5 a_0, 0\;$ in (a) and $r_0=5 a_0, 4 a_0, 3 a_0, 0\;$ in (b). (c) and (d) show the value of $\sigma_{min}$ in monolayer graphene as a function of $r_0/r_i$.
}
\label{fig:MLGemt}
\end{figure}

%
To describe the transport properties close to the CNP and take into
account the strong disorder-induced carrier density inhomogeneities we
use the effective medium theory (EMT), where the
conductivity is found by solving the following integral equation
\cite{bruggeman1935,landauer1952,landauer1978,dassarma2010,Rossi,fogler2009,DasHwangLi_arXiv11}:
\beq
\int dn \dfrac{\sigma(n)-\sigma_{EMT}}{\sigma(n) + \sigma_{EMT}} P(n) = 0
 \label{eq:emt}
\eeq
where $\sigma (n)$ is the local Boltzmann conductivity obtained in
Section \ref{subsec:mlgbtt}. Fig.~\ref{fig:MLGemt}(a) and (b) show
the EMT results for $\sigma(n)$. The EMT results give similar behavior
of $\sigma(n)$ at high carrier density as shown in
Fig.~\ref{fig:MLGcond}, where the density fluctuations are strongly
suppressed. However,  close to the Dirac point, the graphene
conductivity obtained using TFD-EMT approach is approximately a
constant, with this constant minimum conductivity plateau strongly
depending on the correlation length $r_0$. Fig.~\ref{fig:MLGemt}(c)
and (d) show the dependence of $\sigma_{min}$ on the size of the
correlation length $r_0$. $\sigma_{min}$ increases slowly with $r_0$
for $r_0/r_i<0.5$, but quite rapidly for $r_0/r_i>0.5$.
The results in Fig.~\ref{fig:MLGemt}(c) and (d)
are in qualitative agreement with the scaling of $\sigma_{min}$
with temperature, proportional to $r_0$, observed in experiments
\cite{fuhrer2010}.

\section{Bilayer graphene conductivity}
\label{sec:blg}

In this section we extend the theory presented in the previous section
for monolayer graphene to bilayer graphene.
 MLG.
The most important difference between MLG and BLG
comes from the fact  that, in BLG, at low energies, the band dispersion is
approximately parabolic with effective mass $m \simeq
0.033m_e$ ($m_e$ being the bare electron mass) \cite{mccann2006}
rather than linear as in MLG. As a consequence in BLG the scaling
of the conductivity with doping, at high density, differs from
the one in MLG.
We restrict ourselves to the case in which no perpendicular
electric field is present so that no gap is present between
the conduction and the valence band
\cite{castro2007,oostinga2008,mak2009,zhujun_PRBR10,Rossi_PRL11}.

To characterize the spatial correlation among charged impurities
we use the same model that we used for MLG.

\subsection{High density: Boltzmann transport theory}

Within the two-band approximation, the BLG conductivity at zero temperature $T=0$ is given by:
\beq
\sigma  =\dfrac{e^{2} n \tau}{m}
\label{eq:blgsig}
\eeq
where $\tau$ is the relaxation time in BLG for the case in which
the charged impurities are spatially correlated.
$\tau$ is given by Eq. \ref{eq:mscatt}
with
$\epsilon_{s\mathbf{k}} = s \hbar^2 k^2/2m$ for the pseudo-spin state
``$s$", $\epsilon(|\bk-\bk'|)$ the static dielectric screening function of BLG
Ref.~[\onlinecite{HwangBG_PRL08}], and
$g(\theta_{\bf kk'})=\left[1+\cos2\theta_{\bf kk'}\right]/2$
the chiral factor for states on the lowest energy bands of BLG.

The full static dielectric constant of gapless BLG at $T=0$ is given by \cite{HwangBG_PRL08}
%
\begin{equation}
 \begin{array}{l l}
 \varepsilon(q) & = [1+V(q)\Pi(q)]^{-1} \\
             & = [1+V(q)D_0\left[g(q)-f(q)\theta(q-2k_{F})\right]]^{-1}
 \end{array}
\label{eq:fullpo}
\end{equation}
where $\Pi (q)$ is the BLG static polarizability,
$D_0=\dfrac{2 m }{ \pi \hbar^2}$ the density of states, and
\begin{equation}
\begin{array}{l l l }
f(q) & =\dfrac{2k_{F}^{2}+q^{2}}{2k_{F}^{2}q}\sqrt{q^{2}-4k_{F}^{2}}+\ln\dfrac{q-\sqrt{q^{2}-4k_{F}^{2}}}{q+\sqrt{q^{2}-4k_{F}^{2}}}\\
g(q) & =\dfrac{1}{2k_{F}^{2}}\sqrt{q^{4}+4k_{F}^{4}}-\ln\left[\dfrac{k_{F}^{2}+\sqrt{k_{F}^{4}+q^{4}/4}}{2k_{F}^{2}}\right]
\end{array}
\end{equation}
To make analytical progress, we calculate the density-dependent conductivity using the dielectric function of BLG within the Thomas-Fermi approximation:
\begin{equation}
\varepsilon(q)=1+\dfrac{q_{TF}}{q}  
\end{equation}
where $q_{TF}=\dfrac{4 m e^2}{\kappa \hbar^2} \simeq 1.0 \times 10^{9}$m$^{-1}$ for bilayer graphene on $\text{SiO}_{2}$
substrate, which is a density independent constant and is larger than $2 k_F$ for carrier density $n < 8\times 10^{12}$cm$^{-2}$. The relaxation time including correlated disorder is then simplified as:
\beq
\dfrac{\hbar}{\tau}=\dfrac{n_{i}\pi\ensuremath{\hbar^{2}}q_{0}^{2}}{m}\int_{0}^{1}dx\left[\dfrac{1}{x+q_{0}}\right]^{2}\dfrac{x^{2}\left(1-2x^{2}\right)^{2}}{\sqrt{1-x^{2}}}S(2k_{F}x)
\label{eq:blgrelatime}
\eeq
where $q_{0}  =q_{TF}/(2k_{F})$. To incorporate analytically the correlation effects of charged impurities,
we again expand $S (x)$ around $x \sim 0$:
\begin{equation}
S( 2 k_F x)\simeq 1-a+\dfrac{1}{2}\dfrac{n}{n_{i}}a^{2}x^{2}-\dfrac{1}{12}\dfrac{n^{2}}{n_{i}^{2}}a^{3}x^{4}
\label{eq:blgstruc}
\end{equation}

Combining Eqs.~(\ref{eq:blgsig}), (\ref{eq:blgrelatime}), and
(\ref{eq:blgstruc}) we obtain for $\sigma(n)$ at $T=0$ in the presence
of correlated disorder
\begin{equation}
\sigma =\dfrac{e^{2}}{h}\dfrac{2n}{n_{i}}\frac{1}{\left[(1-a)G_{1}[q_{0}]+\frac{n}{2n_{i}}a^{2}G_{2}[q_{0}]-\frac{n^{2}}{12 n_{i}^{2}}a^{3}G_{3}[q_{0}]\right]},
\label{eq:Anasigma}
\end{equation}
where
\begin{equation}
\begin{array}{l l l l l }
G_{1}(q_{0}) & =q_{0}^{2}\int_{0}^{1}\dfrac{1}{(x+q_{0})^{2}}\dfrac{x^{2}\left(1-2x^{2}\right)^{2}}{\sqrt{1-x^{2}}}dx\\
G_{2}(q_{0}) & =q_{0}^{2}\int_{0}^{1}\dfrac{1}{(x+q_{0})^{2}}\dfrac{x^{4}\left(1-2x^{2}\right)^{2}}{\sqrt{1-x^{2}}}dx\\
G_{3}(q_{0}) & =q_{0}^{2}\int_{0}^{1}\dfrac{1}{(x+q_{0})^{2}}\dfrac{x^{6}\left(1-2x^{2}\right)^{2}}{\sqrt{1-x^{2}}}dx
\end{array}
\label{eq:g1}
\end{equation}
For each value of $r_0$ and carrier density $n$, the resistivity of BLG for correlated disorder is also not a linear function of impurity density, and its behavior is close to that in MLG. The maximum resistivity of BLG is found to be at
\begin{equation}
r_i/r_0 = \sqrt{2(1-\pi B_B \pi n r_0^2- C_B \pi^2 n^2 r_0^4)}.
\label{eq:optimalBLG}
\end{equation}
with $B_B = G_2[q_0]/(2G_1[q_0])$ and $C_B=-G_3[q_0]/(12G_1[q_0])$, which are functions weakly depending on carrier density $n$.

It is straightforward to calculate the asymptotic density dependence of BLG conductivity from the above formula and we will discuss $\sigma(n)$ in the strong ($q_0 \gg 1$) and weak $q_0 \ll 1$ screening limits separately.

In the strong screening limit $q_0 \gg 1$, $G_{1}[q_0] \simeq \pi/8$, $G_{2}[q_0]\simeq 7 \pi/64$  and $G_{3}[q_0]\simeq 13 \pi/128$. For randomly distributed charged impurity,  we can express the conductivity as a linear function of carrier density $\sigma(n) \sim n$ \cite{DasEnrico_PRB10}. In the presence of correlated charged impurity
we find:
\begin{equation}
\sigma(n) = \dfrac{A_B n}{1- a + a^2 \dfrac{7 n}{16n_{i}}+ a^3 \dfrac{13 n^2}{192 n^2_{i}}},
\label{eq:fisigma}
\end{equation}
where $a= \pi n_{i} r_0^2$, and $A_B \simeq \dfrac{e^{2}}{h}\dfrac{16}{\pi n_{i}}$.
In the strong screening limit $q_0 \gg 1\Rightarrow n\ll n_i$
from \ceq{eq:fisigma} we obtain $\sigma(n)\sim  A_B n/(1-a)$.
With the increase of carrier density, the calculated conductivity in BLG also shows the sublinear behavior as in MLG due to the third and fourth terms in the denominator of Eq. \ref{eq:fisigma}.

In the weak screening limit, $q_0 \ll 1$, we have $G_{1}[q_0] \simeq \pi q_{0}^{2}/4$, $G_{2}[q_0] \simeq \pi q_{0}^{2}/8$ and $G_{3}[q_0] \simeq 7\pi q_{0}^{2}/64$. The conductivity of BLG in the limit $q_0 \ll 1$ is a quadratic function of carrier density for randomly distributed Coulomb disorder:
\begin{equation}
\sigma(n) = \dfrac{e^{2}}{h}\dfrac{32n^2}{ n_{i}q^2_{TF}}
\end{equation}
For the correlated disorder, the calculated conductivity of BLG shows the sub-quadratic behavior:
\begin{equation}
\sigma(n) = \dfrac{A_b n^2}{1- a + a^2 \dfrac{n}{4n_{i}}- a^3 \dfrac{7n^2}{192n^2_{i}}},
\label{eq:fisigma2}
\end{equation}
with $A_b = \dfrac{e^{2}}{h}\dfrac{32}{ n_{i}q^2_{TF}}$.

\begin{figure}
\includegraphics[width=0.99\columnwidth]{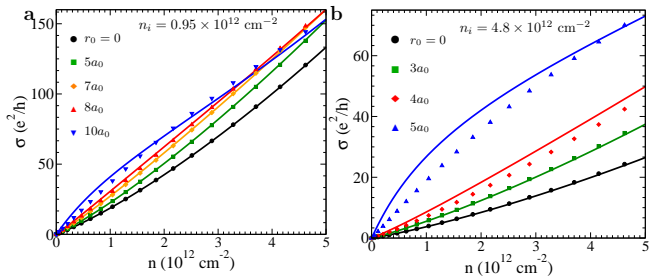}
\caption{
Calculated $\sigma(n)$ in bilayer graphene with $S(\bq)$ obtained
  from the Monte Carlo
  simulations (symbols) and $S(\bq)$ given by Eq.~(\ref{eq:strufac2}) (solid lines) for
  two different impurity densities (a) $n_i=0.95\times 10^{12}$
  cm$^{-2}$ and (b) $n_i=4.8\times
  10^{12}$ cm$^{-2}$. The different lines correspond to different
  values of $r_0$. In (a) we use $r_0=10 a_0, \; 8
  a_0, \; 7 a_0, \; 5 a_0, \; 0$ (from top to bottom),  and in (b)
  $r_0=5 a_0, \; 4 a_0, \; 3 a_0, \; 0 $ (from top to bottom).
}
\label{fig:BLGcond}
\end{figure}

In Figs.~\ref{fig:BLGcond}(a) and (b), we show the $\sigma (n)$
within Boltzmann transport theory obtained numerically taking into
account the screening via the static dielectric function
given by Eq. \ref{eq:fullpo}. We show the results
for several different correlation
lengths $r_0$ and two different charged impurity densities, (a)
$n_i=0.95 \times 10^{12}$ cm$^{-2}$ and (b) $n_i = 4.8\times 10^{12}$
cm$^{-2}$.
From Figs.~\ref{fig:BLGcond}(a),~(b) we see that the conductivity increases
with $r_0$ as in MLG.
However the details of the scaling of $\sigma$ with doping differ between
MLG and BLG. In BLG $\sigma(n)\approx n^\alpha$ where $1<\alpha<2$ also
depends on $n$.
The effect of spatial correlations among impurities
in BLG is to increase $\alpha$ at low densities and reduce it at high densities.

\begin{figure}
\includegraphics[width=0.99\columnwidth]{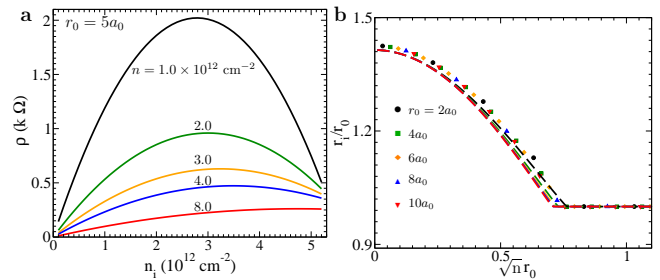}
\caption{
(a) The resistivity $\rho$ in bilayer graphene is shown as a function of impurity density $n_i$ for different carrier densities
  with $r_0 =5 a_0$. (b) The relationship between $r_i/r_0$
  and $\sqrt{n} r_0$ in bilayer graphene, where the conductivity is minimum. The dashed lines are obtained using Eq. \ref{eq:optimalBLG}.
}
\label{fig:Fig8_BLGoptimal}
\end{figure}

In Fig.~\ref{fig:Fig8_BLGoptimal}(a), we present the resistivity of
BLG as a function of impurity density for various carrier density with
$r_0 = 5a_0$. The spatial correlation of charged impurity leads to
a highly non-linear function of $\rho (n_i)$ as in MLG. We also
present the relation between $r_i/r_0$ and $\sqrt{n}r_0$ where the
maximum resistivity of BLG occurs in
Fig.~\ref{fig:Fig8_BLGoptimal}(b). The results are quite close to
those of MLG shown in Fig.~\ref{fig:MLGoptimal}.

\subsection{Low density: Effective medium theory}
As in MLG, also in BLG, because of the gapless nature of the
dispersion the presence of charged impurities induces large
carrier density fluctuations \cite{adam2008a,deshpande2009b,DasEnrico_PRB10,Rossi_PRL11}
that strongly affect the transport properties of BLG.

\begin{figure}[tb]
 \begin{center}
  \includegraphics[width=8.5cm]{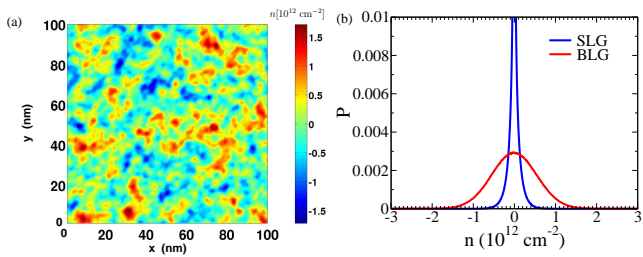}
  \caption{
           (Color online).
           (a) $n(\br)$ of BLG at the CNP
           for a single
           disorder realization with $n_i=10^{11}{\rm cm}^{-2}$ and $d=1$~nm.
           (b) Disorder averaged $P(n)$,
           at the CNP for BLG (MLG) red (blue) for
           $n_i=10^{11}{\rm cm}^{-2}$ and $d=1$~nm.
           For MLG $P(n=0)\approx 0.1$, out of scale.
           The corresponding $n_{\rm rms}$ is $5.5 \times 10^{11}{\rm
           cm}^{-2}$ for BLG
           and $1.2 \times 10^{11}{\rm cm}^{-2}$ for MLG.
          }
  \label{fig:blg_n}
 \end{center}
\end{figure}

Fig.~\ref{fig:blg_n}(a) shows the calculated density landscape for
BLG for a single disorder realization, and Fig.~\ref{fig:blg_n}(a)
a comparison of the probability distribution function $P(n)$
for BLG and MLG \cite{DasEnrico_PRB10}.
%
Within the Thomas-Fermi approximation, approximating the low energy
bands as parabolic, in BLG, with no spatial correlation between
charged impurities, $P(n)$ is a Gaussian whose root mean
square is independent of the doping and is
given by the following equation \cite{Rossi_PRL11}:
\beq
 \nrms        = \frac{\sqrt{n_i}}{\rb}\left[\frac{2}{\pi}f(d/\rb)\right]^{1/2}
 \label{eq:blg_nrms}
\enq
where $f(d/\rb) = e^{2d/\rb}(1+2d/\rb)\Gamma(0,2d/\rb) - 1$
is a dimensionless function,
$\rb \equiv [(2e^2m^*)/(\kappa\hbar^2)]^{-1}\approx 2$~nm is the screening length,
and $\Gamma(a,x)$ is the incomplete gamma function.
For small $d/\rb$, $f=-1-\gamma-\log(2d/\rb)+O(d/\rb)$ (where $\gamma=0.577216$ is the Euler constant), whereas for $d\gg \rb$
$f=1/(2d/\rb)^2 + O((d/\rb)^{-3})$.
As for MLG, also for BLG we find that the presence of spatial correlations among impurities
has only a minor quantitative effect on $P(n)$. For this reason, and the fact that
with no correlation between the impurities, $P(n)$ has a particularly simple analytical
expression, for BLG we neglect the effect of impurity spatial correlations on $P(n)$.

\begin{figure}
 \includegraphics[width=0.99\columnwidth]{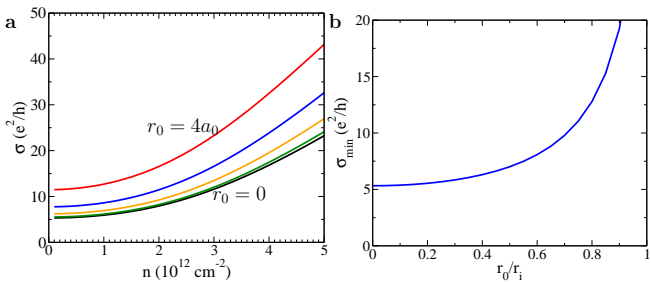}
\caption{
  (a) BLG conductivity as a function of $n$ obtained using the EMT for
  $n_i=4.8 \times 10^{12}$ cm$^{-2}$ for
  $r_0 = (4,3,2,1,0)\times a_0$ from top to bottom.
  (b) BLG $\sigma_{\rm min}$ as a function of $r_0/r_i$ for  $n_i=4.8 \times 10^{12}$ cm$^{-2}$.
}
\label{fig:BLGcondEMT}
\end{figure}

As in MLG the effect  of the strong carrier density inhomogeneities on transport
can be effectively taken into account using the effective medium theory.
Using Eq.~\ceq{eq:emt}, $\sigma(n)$ given by the Boltzmann theory,
and $P(n)$ as described in the previous paragraph, the effective conductivity
$\sigma_{EMT}$ for BLG can be calculated taking into account the presence of strong
carrier density fluctuations.
Fig.~\ref{fig:BLGcondEMT}(a) shows the scaling of $\sigma$ with doping obtained
using the EMT for several values of $r_0$ and
$n_i=4.8\times   10^{12}$ cm$^{-2}$.
Taking account of the carrier density inhomogeneities that dominates close
to the charge neutrality point, the EMT returns a non-zero value of the conductivity $\sigma_{\rm min}$
for zero average density, a value that depends on the impurity density
and their spatial correlations. In particular, as shown in Fig.~\ref{fig:BLGcondEMT}(b),
in analogy to the MLG case $\sigma_{\rm min}$ grows with $r_0$.

\section{Discussion of experiments}
\label{sec:discuss}
Although the sublinearity of $\sigma(n)$ can be explained by including
both long- and short-range scatterers (or resonant scatterers) in the
Boltzmann transport theory \cite{DasHwang_BN11}, it can not explain
the observed enhancement of conductivity with increasing annealing
temperatures as observed in Ref.~[\onlinecite{fuhrer2010}].
Annealing leads to
stronger correlations among the impurities since the impurities can
move around to equilibrium sites.
Our results show that by increasing $r_0$, at low densities,
both the conductivity and the mobility of MLG and BLG increase.
Moreover, our results for MLG \cite{qzli_PRL11} show that as $r_0$
increases the crossover density at which $\sigma(n)$ from linear
becomes sublinear decreases.
All these features have been observed experimentally for MLG \cite{fuhrer2010}.
In addition, our transport
theory based on the correlated impurity model also gives a possible
explanation for the observed strong nonlinear $\sigma(n)$ in
suspended graphene \cite{BolotinYacoby,Feldman2} where the
thermal/current annealing is used routinely.
No experiment has so far directly studied the effect of increasing the
spatial correlations among charged impurities in BLG and tested our predictions for BLG.

Although we have used a minimal model for
impurity correlations, using a single correlation length parameter
$r_0$, which captures the essential physics of correlated impurity
scattering, it should be straightforward to improve the model with
more sophisticated correlation models if experimental information on
impurity correlations becomes available \cite{fuhrer2010}. Intentional
control of spatial charged impurity distributions or by rapid thermal
annealing and quenching, should be a powerful tool to further increase
mobility in monolayer and bilayer graphene devices\cite{fuhrer2010}.

\section{Conclusions}
\label{sec:conclu}
In summary, we provide a novel physically motivated explanation for
the observed sublinear scaling of the graphene conductivity with density
at high dopings
by showing that the inclusion of spatial
correlations among the charged impurity locations leads to a
significant sublinear density dependence in the conductivity of MLG in contrast to the strictly linear-in-density graphene conductivity for
uncorrelated random charged impurity scattering. We also show that the spatial correlation of charged impurity will also enhance the mobility of BLG. The great merit of
our theory is that it  eliminates the need for an {\it ad hoc}
zero-range defect scattering mechanism which has always been used in
the standard model of graphene transport in order to
phenomenologically explain the high-density sublinear behavior $\sigma(n)$ of MLG. Even though the
short-range disorder is not needed to explain the sublinear behavior of $\sigma(n)$
in our model we do not exclude the possibility of short range
disorder scattering in real MLG samples, which would just add as
another resistive channel with constant resistivity.
Our theoretical results are confirmed qualitatively
by the experimental measurements presented in Ref. [\onlinecite{fuhrer2010}]
in which the spatial correlations among charged impurities were modified
via thermal annealing with no change of the impurity density.
Our results, combined with the experimental observation of Ref. [\onlinecite{fuhrer2010}],
demonstrate that in monolayer and bilayer graphene samples in which
charged impurities are the dominant source of scattering the mobility
can be greatly enhanced by thermal/current annealing processes that
increase the spatial correlations among the impurities.

\section{Acknowledgements}

This work is supported by ONR-MURI and NRI-SWAN.
ER acknowledges support from the Jeffress Memorial Trust, Grant No. J-1033.
ER and EHH acknowledge the hospitality of KITP, supported in part
by the National Science Foundation under Grant No. PHY11-25915,
where part of this work was done.
Computations were carried out in part on the SciClone Cluster at the College of William and Mary.


\begin{thebibliography}{60}
\expandafter\ifx\csname natexlab\endcsname\relax\def\natexlab#1{#1}\fi
\expandafter\ifx\csname bibnamefont\endcsname\relax
  \def\bibnamefont#1{#1}\fi
\expandafter\ifx\csname bibfnamefont\endcsname\relax
  \def\bibfnamefont#1{#1}\fi
\expandafter\ifx\csname citenamefont\endcsname\relax
  \def\citenamefont#1{#1}\fi
\expandafter\ifx\csname url\endcsname\relax
  \def\url#1{\texttt{#1}}\fi
\expandafter\ifx\csname urlprefix\endcsname\relax\def\urlprefix{URL }\fi
\providecommand{\bibinfo}[2]{#2}
\providecommand{\eprint}[2][]{\url{#2}}

\bibitem[{\citenamefont{Novoselov et~al.}(2004)\citenamefont{Novoselov, Geim,
  Morozov, Jiang, Zhang, Dubonos, Grigorieva, and Firsov}}]{novoselov2004}
\bibinfo{author}{\bibfnamefont{K.~S.} \bibnamefont{Novoselov}},
  \bibinfo{author}{\bibfnamefont{A.~K.} \bibnamefont{Geim}},
  \bibinfo{author}{\bibfnamefont{S.~V.} \bibnamefont{Morozov}},
  \bibinfo{author}{\bibfnamefont{D.}~\bibnamefont{Jiang}},
  \bibinfo{author}{\bibfnamefont{Y.}~\bibnamefont{Zhang}},
  \bibinfo{author}{\bibfnamefont{S.~V.} \bibnamefont{Dubonos}},
  \bibinfo{author}{\bibfnamefont{I.~V.} \bibnamefont{Grigorieva}},
  \bibnamefont{and} \bibinfo{author}{\bibfnamefont{A.~A.}
  \bibnamefont{Firsov}}, \bibinfo{journal}{Science}
  \textbf{\bibinfo{volume}{306}}, \bibinfo{pages}{666} (\bibinfo{year}{2004}).

\bibitem[{\citenamefont{Das~Sarma et~al.}(2011)\citenamefont{Das~Sarma, Adam,
  Hwang, and Rossi}}]{dassarma2010}
\bibinfo{author}{\bibfnamefont{S.}~\bibnamefont{Das~Sarma}},
  \bibinfo{author}{\bibfnamefont{S.}~\bibnamefont{Adam}},
  \bibinfo{author}{\bibfnamefont{E.~H.} \bibnamefont{Hwang}}, \bibnamefont{and}
  \bibinfo{author}{\bibfnamefont{E.}~\bibnamefont{Rossi}},
  \bibinfo{journal}{Rev. Mod. Phys.} \textbf{\bibinfo{volume}{83}},
  \bibinfo{pages}{407} (\bibinfo{year}{2011}).

\bibitem[{\citenamefont{Peres}(2010)}]{peres_RMP10}
\bibinfo{author}{\bibfnamefont{N.~M.~R.} \bibnamefont{Peres}},
  \bibinfo{journal}{Rev. Mod. Phys.} \textbf{\bibinfo{volume}{82}},
  \bibinfo{pages}{2673} (\bibinfo{year}{2010}).

\bibitem[{\citenamefont{Chen et~al.}(2009)\citenamefont{Chen, Cullen, Jang,
  Fuhrer, and Williams}}]{FuhrerDefect_PRL}
\bibinfo{author}{\bibfnamefont{J.-H.} \bibnamefont{Chen}},
  \bibinfo{author}{\bibfnamefont{W.~G.} \bibnamefont{Cullen}},
  \bibinfo{author}{\bibfnamefont{C.}~\bibnamefont{Jang}},
  \bibinfo{author}{\bibfnamefont{M.~S.} \bibnamefont{Fuhrer}},
  \bibnamefont{and} \bibinfo{author}{\bibfnamefont{E.~D.}
  \bibnamefont{Williams}}, \bibinfo{journal}{Phys. Rev. Lett.}
  \textbf{\bibinfo{volume}{102}}, \bibinfo{pages}{236805}
  (\bibinfo{year}{2009}).

\bibitem[{\citenamefont{Ishigami et~al.}(2007)\citenamefont{Ishigami, Chen,
  Cullen, Fuhrer, and Williams}}]{Ishigami_Nano07}
\bibinfo{author}{\bibfnamefont{M.}~\bibnamefont{Ishigami}},
  \bibinfo{author}{\bibfnamefont{J.~H.} \bibnamefont{Chen}},
  \bibinfo{author}{\bibfnamefont{W.~G.} \bibnamefont{Cullen}},
  \bibinfo{author}{\bibfnamefont{M.~S.} \bibnamefont{Fuhrer}},
  \bibnamefont{and} \bibinfo{author}{\bibfnamefont{E.~D.}
  \bibnamefont{Williams}}, \bibinfo{journal}{Nano Letters}
  \textbf{\bibinfo{volume}{7}}, \bibinfo{pages}{1643} (\bibinfo{year}{2007}).

\bibitem[{\citenamefont{Katsnelson and Geim}(2008)}]{katsnelson2008}
\bibinfo{author}{\bibfnamefont{M.~I.} \bibnamefont{Katsnelson}}
  \bibnamefont{and} \bibinfo{author}{\bibfnamefont{A.~K.} \bibnamefont{Geim}},
  \bibinfo{journal}{Phil. Trans. R. Soc. A} \textbf{\bibinfo{volume}{366}},
  \bibinfo{pages}{195} (\bibinfo{year}{2008}).

\bibitem[{\citenamefont{Bao et~al.}(2009)\citenamefont{Bao, Miao, Chen, Zhang,
  Jang, Dames, and Lau}}]{Bao_NatTech09}
\bibinfo{author}{\bibfnamefont{W.}~\bibnamefont{Bao}},
  \bibinfo{author}{\bibfnamefont{F.}~\bibnamefont{Miao}},
  \bibinfo{author}{\bibfnamefont{Z.}~\bibnamefont{Chen}},
  \bibinfo{author}{\bibfnamefont{H.}~\bibnamefont{Zhang}},
  \bibinfo{author}{\bibfnamefont{W.}~\bibnamefont{Jang}},
  \bibinfo{author}{\bibfnamefont{C.}~\bibnamefont{Dames}}, \bibnamefont{and}
  \bibinfo{author}{\bibfnamefont{C.~N.} \bibnamefont{Lau}},
  \bibinfo{journal}{Nature Nanotech.} \textbf{\bibinfo{volume}{4}},
  \bibinfo{pages}{562} (\bibinfo{year}{2009}).

\bibitem[{\citenamefont{Stauber et~al.}(2007)\citenamefont{Stauber, Peres, and
  Guinea}}]{Stauber_resonantPRB07}
\bibinfo{author}{\bibfnamefont{T.}~\bibnamefont{Stauber}},
  \bibinfo{author}{\bibfnamefont{N.~M.~R.} \bibnamefont{Peres}},
  \bibnamefont{and} \bibinfo{author}{\bibfnamefont{F.}~\bibnamefont{Guinea}},
  \bibinfo{journal}{Phys. Rev. B} \textbf{\bibinfo{volume}{76}},
  \bibinfo{pages}{205423} (\bibinfo{year}{2007}).

\bibitem[{\citenamefont{Monteverde et~al.}(2010)\citenamefont{Monteverde,
  Ojeda-Aristizabal, Weil, Bennaceur, Ferrier, Gu\'eron, Glattli, Bouchiat,
  Fuchs, and Maslov}}]{Monteverde_PRL10}
\bibinfo{author}{\bibfnamefont{M.}~\bibnamefont{Monteverde}},
  \bibinfo{author}{\bibfnamefont{C.}~\bibnamefont{Ojeda-Aristizabal}},
  \bibinfo{author}{\bibfnamefont{R.}~\bibnamefont{Weil}},
  \bibinfo{author}{\bibfnamefont{K.}~\bibnamefont{Bennaceur}},
  \bibinfo{author}{\bibfnamefont{M.}~\bibnamefont{Ferrier}},
  \bibinfo{author}{\bibfnamefont{S.}~\bibnamefont{Gu\'eron}},
  \bibinfo{author}{\bibfnamefont{C.}~\bibnamefont{Glattli}},
  \bibinfo{author}{\bibfnamefont{H.}~\bibnamefont{Bouchiat}},
  \bibinfo{author}{\bibfnamefont{J.~N.} \bibnamefont{Fuchs}}, \bibnamefont{and}
  \bibinfo{author}{\bibfnamefont{D.~L.} \bibnamefont{Maslov}},
  \bibinfo{journal}{Phys. Rev. Lett.} \textbf{\bibinfo{volume}{104}},
  \bibinfo{pages}{126801} (\bibinfo{year}{2010}).

\bibitem[{\citenamefont{Wehling et~al.}(2010)\citenamefont{Wehling, Yuan,
  Lichtenstein, Geim, and Katsnelson}}]{wehiling2010c}
\bibinfo{author}{\bibfnamefont{T.~O.} \bibnamefont{Wehling}},
  \bibinfo{author}{\bibfnamefont{S.}~\bibnamefont{Yuan}},
  \bibinfo{author}{\bibfnamefont{A.~I.} \bibnamefont{Lichtenstein}},
  \bibinfo{author}{\bibfnamefont{A.~K.} \bibnamefont{Geim}}, \bibnamefont{and}
  \bibinfo{author}{\bibfnamefont{M.~I.} \bibnamefont{Katsnelson}},
  \bibinfo{journal}{Phys. Rev. Lett.} \textbf{\bibinfo{volume}{105}},
  \bibinfo{pages}{056802} (\bibinfo{year}{2010}).

\bibitem[{\citenamefont{Ferreira et~al.}(2011)\citenamefont{Ferreira,
  Viana-Gomes, Nilsson, Mucciolo, Peres, and Castro~Neto}}]{Ferreira_PRB11}
\bibinfo{author}{\bibfnamefont{A.}~\bibnamefont{Ferreira}},
  \bibinfo{author}{\bibfnamefont{J.}~\bibnamefont{Viana-Gomes}},
  \bibinfo{author}{\bibfnamefont{J.}~\bibnamefont{Nilsson}},
  \bibinfo{author}{\bibfnamefont{E.~R.} \bibnamefont{Mucciolo}},
  \bibinfo{author}{\bibfnamefont{N.~M.~R.} \bibnamefont{Peres}},
  \bibnamefont{and} \bibinfo{author}{\bibfnamefont{A.~H.}
  \bibnamefont{Castro~Neto}}, \bibinfo{journal}{Phys. Rev. B}
  \textbf{\bibinfo{volume}{83}}, \bibinfo{pages}{165402}
  (\bibinfo{year}{2011}).

\bibitem[{\citenamefont{Efetov and Kim}(2010)}]{EfetovKim_PRL10}
\bibinfo{author}{\bibfnamefont{D.~K.} \bibnamefont{Efetov}} \bibnamefont{and}
  \bibinfo{author}{\bibfnamefont{P.}~\bibnamefont{Kim}},
  \bibinfo{journal}{Phys. Rev. Lett.} \textbf{\bibinfo{volume}{105}},
  \bibinfo{pages}{256805} (\bibinfo{year}{2010}).

\bibitem[{\citenamefont{Hwang and
  Das~Sarma}(2008{\natexlab{a}})}]{HwangDasPhonon_PRB08}
\bibinfo{author}{\bibfnamefont{E.~H.} \bibnamefont{Hwang}} \bibnamefont{and}
  \bibinfo{author}{\bibfnamefont{S.}~\bibnamefont{Das~Sarma}},
  \bibinfo{journal}{Phys. Rev. B} \textbf{\bibinfo{volume}{77}},
  \bibinfo{pages}{115449} (\bibinfo{year}{2008}{\natexlab{a}}).

\bibitem[{\citenamefont{Min et~al.}(2011)\citenamefont{Min, Hwang, and
  Das~Sarma}}]{MinHwangDas_arX10}
\bibinfo{author}{\bibfnamefont{H.}~\bibnamefont{Min}},
  \bibinfo{author}{\bibfnamefont{E.~H.} \bibnamefont{Hwang}}, \bibnamefont{and}
  \bibinfo{author}{\bibfnamefont{S.}~\bibnamefont{Das~Sarma}},
  \bibinfo{journal}{Phys. Rev. B} \textbf{\bibinfo{volume}{83}},
  \bibinfo{pages}{161404} (\bibinfo{year}{2011}).

\bibitem[{\citenamefont{Li et~al.}(2011{\natexlab{a}})\citenamefont{Li, Hwang,
  and Das~Sarma}}]{qzli_PRB11}
\bibinfo{author}{\bibfnamefont{Q.}~\bibnamefont{Li}},
  \bibinfo{author}{\bibfnamefont{E.~H.} \bibnamefont{Hwang}}, \bibnamefont{and}
  \bibinfo{author}{\bibfnamefont{S.}~\bibnamefont{Das~Sarma}},
  \bibinfo{journal}{Phys. Rev. B} \textbf{\bibinfo{volume}{84}},
  \bibinfo{pages}{115442} (\bibinfo{year}{2011}{\natexlab{a}}).

\bibitem[{\citenamefont{Heo et~al.}(2011)\citenamefont{Heo, Chung, Lee, Yang,
  Seo, Shin, Chung, Seo, Hwang, and Das~Sarma}}]{Heo_PRB11}
\bibinfo{author}{\bibfnamefont{J.}~\bibnamefont{Heo}},
  \bibinfo{author}{\bibfnamefont{H.~J.} \bibnamefont{Chung}},
  \bibinfo{author}{\bibfnamefont{S.-H.} \bibnamefont{Lee}},
  \bibinfo{author}{\bibfnamefont{H.}~\bibnamefont{Yang}},
  \bibinfo{author}{\bibfnamefont{D.~H.} \bibnamefont{Seo}},
  \bibinfo{author}{\bibfnamefont{J.~K.} \bibnamefont{Shin}},
  \bibinfo{author}{\bibfnamefont{U.-I.} \bibnamefont{Chung}},
  \bibinfo{author}{\bibfnamefont{S.}~\bibnamefont{Seo}},
  \bibinfo{author}{\bibfnamefont{E.~H.} \bibnamefont{Hwang}}, \bibnamefont{and}
  \bibinfo{author}{\bibfnamefont{S.}~\bibnamefont{Das~Sarma}},
  \bibinfo{journal}{Phys. Rev. B} \textbf{\bibinfo{volume}{84}},
  \bibinfo{pages}{035421} (\bibinfo{year}{2011}).

\bibitem[{\citenamefont{Hwang and
  Das~Sarma}(2008{\natexlab{b}})}]{HwangDasGaAsmobi_PRB08}
\bibinfo{author}{\bibfnamefont{E.~H.} \bibnamefont{Hwang}} \bibnamefont{and}
  \bibinfo{author}{\bibfnamefont{S.}~\bibnamefont{Das~Sarma}},
  \bibinfo{journal}{Phys. Rev. B} \textbf{\bibinfo{volume}{77}},
  \bibinfo{pages}{235437} (\bibinfo{year}{2008}{\natexlab{b}}).

\bibitem[{\citenamefont{Novoselov
  et~al.}(2005{\natexlab{a}})\citenamefont{Novoselov, Geim, Morozov, Jiang,
  Katsnelson, Grigorieva, Dubonos, and Firsov}}]{Novoselov2}
\bibinfo{author}{\bibfnamefont{K.~S.} \bibnamefont{Novoselov}},
  \bibinfo{author}{\bibfnamefont{A.~K.} \bibnamefont{Geim}},
  \bibinfo{author}{\bibfnamefont{S.~V.} \bibnamefont{Morozov}},
  \bibinfo{author}{\bibfnamefont{D.}~\bibnamefont{Jiang}},
  \bibinfo{author}{\bibfnamefont{M.~I.} \bibnamefont{Katsnelson}},
  \bibinfo{author}{\bibfnamefont{I.~V.} \bibnamefont{Grigorieva}},
  \bibinfo{author}{\bibfnamefont{S.~V.} \bibnamefont{Dubonos}},
  \bibnamefont{and} \bibinfo{author}{\bibfnamefont{A.~A.}
  \bibnamefont{Firsov}}, \bibinfo{journal}{Nature}
  \textbf{\bibinfo{volume}{438}}, \bibinfo{pages}{197}
  (\bibinfo{year}{2005}{\natexlab{a}}).

\bibitem[{\citenamefont{Tan et~al.}(2007)\citenamefont{Tan, Zhang, Bolotin,
  Zhao, Adam, Hwang, Das~Sarma, Stormer, and Kim}}]{TanDas_PRL07}
\bibinfo{author}{\bibfnamefont{Y.-W.} \bibnamefont{Tan}},
  \bibinfo{author}{\bibfnamefont{Y.}~\bibnamefont{Zhang}},
  \bibinfo{author}{\bibfnamefont{K.}~\bibnamefont{Bolotin}},
  \bibinfo{author}{\bibfnamefont{Y.}~\bibnamefont{Zhao}},
  \bibinfo{author}{\bibfnamefont{S.}~\bibnamefont{Adam}},
  \bibinfo{author}{\bibfnamefont{E.~H.} \bibnamefont{Hwang}},
  \bibinfo{author}{\bibfnamefont{S.}~\bibnamefont{Das~Sarma}},
  \bibinfo{author}{\bibfnamefont{H.~L.} \bibnamefont{Stormer}},
  \bibnamefont{and} \bibinfo{author}{\bibfnamefont{P.}~\bibnamefont{Kim}},
  \bibinfo{journal}{Phys. Rev. Lett.} \textbf{\bibinfo{volume}{99}},
  \bibinfo{pages}{246803} (\bibinfo{year}{2007}).

\bibitem[{\citenamefont{Chen et~al.}(2008)\citenamefont{Chen, Jang, Adam,
  Fuhrer, Williams, and Ishigami}}]{ChenJ_NPH_2007}
\bibinfo{author}{\bibfnamefont{J.-H.} \bibnamefont{Chen}},
  \bibinfo{author}{\bibfnamefont{C.}~\bibnamefont{Jang}},
  \bibinfo{author}{\bibfnamefont{S.}~\bibnamefont{Adam}},
  \bibinfo{author}{\bibfnamefont{M.~S.} \bibnamefont{Fuhrer}},
  \bibinfo{author}{\bibfnamefont{E.~D.} \bibnamefont{Williams}},
  \bibnamefont{and} \bibinfo{author}{\bibfnamefont{M.}~\bibnamefont{Ishigami}},
  \bibinfo{journal}{Nature Phys.} \textbf{\bibinfo{volume}{4}},
  \bibinfo{pages}{377} (\bibinfo{year}{2008}).

\bibitem[{\citenamefont{Bolotin et~al.}(2008)\citenamefont{Bolotin, Sikes,
  Jiang, Klima, Fudenberg, Hone, Kim, and Stormer}}]{BolotinYacoby}
\bibinfo{author}{\bibfnamefont{K.}~\bibnamefont{Bolotin}},
  \bibinfo{author}{\bibfnamefont{K.}~\bibnamefont{Sikes}},
  \bibinfo{author}{\bibfnamefont{Z.}~\bibnamefont{Jiang}},
  \bibinfo{author}{\bibfnamefont{M.}~\bibnamefont{Klima}},
  \bibinfo{author}{\bibfnamefont{G.}~\bibnamefont{Fudenberg}},
  \bibinfo{author}{\bibfnamefont{J.}~\bibnamefont{Hone}},
  \bibinfo{author}{\bibfnamefont{P.}~\bibnamefont{Kim}}, \bibnamefont{and}
  \bibinfo{author}{\bibfnamefont{H.}~\bibnamefont{Stormer}},
  \bibinfo{journal}{Solid State Commun.} \textbf{\bibinfo{volume}{146}},
  \bibinfo{pages}{351} (\bibinfo{year}{2008}).

\bibitem[{\citenamefont{Feldman et~al.}(2009)\citenamefont{Feldman, Martin, and
  Yacoby}}]{Feldman2}
\bibinfo{author}{\bibfnamefont{B.~E.} \bibnamefont{Feldman}},
  \bibinfo{author}{\bibfnamefont{J.}~\bibnamefont{Martin}}, \bibnamefont{and}
  \bibinfo{author}{\bibfnamefont{A.}~\bibnamefont{Yacoby}},
  \bibinfo{journal}{Nature Phys.} \textbf{\bibinfo{volume}{5}},
  \bibinfo{pages}{889} (\bibinfo{year}{2009}).

\bibitem[{\citenamefont{Novoselov
  et~al.}(2005{\natexlab{b}})\citenamefont{Novoselov, Jiang, Schedin, Booth,
  Khotkevich, Morozov, and Geim}}]{Novoselov}
\bibinfo{author}{\bibfnamefont{K.~S.} \bibnamefont{Novoselov}},
  \bibinfo{author}{\bibfnamefont{D.}~\bibnamefont{Jiang}},
  \bibinfo{author}{\bibfnamefont{F.}~\bibnamefont{Schedin}},
  \bibinfo{author}{\bibfnamefont{T.~J.} \bibnamefont{Booth}},
  \bibinfo{author}{\bibfnamefont{V.~V.} \bibnamefont{Khotkevich}},
  \bibinfo{author}{\bibfnamefont{S.~V.} \bibnamefont{Morozov}},
  \bibnamefont{and} \bibinfo{author}{\bibfnamefont{A.~K.} \bibnamefont{Geim}},
  \bibinfo{journal}{Proc.\ Natl.\ Acad.\ Sci.\ USA}
  \textbf{\bibinfo{volume}{102}}, \bibinfo{pages}{10451}
  (\bibinfo{year}{2005}{\natexlab{b}}).

\bibitem[{\citenamefont{Hong et~al.}(2009)\citenamefont{Hong, Zou, and
  Zhu}}]{ZhuExp_PRB09}
\bibinfo{author}{\bibfnamefont{X.}~\bibnamefont{Hong}},
  \bibinfo{author}{\bibfnamefont{K.}~\bibnamefont{Zou}}, \bibnamefont{and}
  \bibinfo{author}{\bibfnamefont{J.}~\bibnamefont{Zhu}},
  \bibinfo{journal}{Phys. Rev. B} \textbf{\bibinfo{volume}{80}},
  \bibinfo{pages}{241415} (\bibinfo{year}{2009}).

\bibitem[{\citenamefont{Adam et~al.}(2007)\citenamefont{Adam, Hwang, Galitski,
  and Sarma}}]{Adam}
\bibinfo{author}{\bibfnamefont{S.}~\bibnamefont{Adam}},
  \bibinfo{author}{\bibfnamefont{E.~H.} \bibnamefont{Hwang}},
  \bibinfo{author}{\bibfnamefont{V.~M.} \bibnamefont{Galitski}},
  \bibnamefont{and} \bibinfo{author}{\bibfnamefont{S.~D.} \bibnamefont{Sarma}},
  \bibinfo{journal}{Proc.\ Natl.\ Acad.\ Sci.\ USA}
  \textbf{\bibinfo{volume}{104}}, \bibinfo{pages}{18392}
  (\bibinfo{year}{2007}).

\bibitem[{\citenamefont{Rossi et~al.}(2009)\citenamefont{Rossi, Adam, and
  Das~Sarma}}]{Rossi}
\bibinfo{author}{\bibfnamefont{E.}~\bibnamefont{Rossi}},
  \bibinfo{author}{\bibfnamefont{S.}~\bibnamefont{Adam}}, \bibnamefont{and}
  \bibinfo{author}{\bibfnamefont{S.}~\bibnamefont{Das~Sarma}},
  \bibinfo{journal}{Phys. Rev. B} \textbf{\bibinfo{volume}{79}},
  \bibinfo{pages}{245423} (\bibinfo{year}{2009}).

\bibitem[{\citenamefont{Hwang et~al.}(2007)\citenamefont{Hwang, Adam, and
  Das~Sarma}}]{HwangAdamDas_PRL07}
\bibinfo{author}{\bibfnamefont{E.~H.} \bibnamefont{Hwang}},
  \bibinfo{author}{\bibfnamefont{S.}~\bibnamefont{Adam}}, \bibnamefont{and}
  \bibinfo{author}{\bibfnamefont{S.}~\bibnamefont{Das~Sarma}},
  \bibinfo{journal}{Phys. Rev. Lett.} \textbf{\bibinfo{volume}{98}},
  \bibinfo{pages}{186806} (\bibinfo{year}{2007}).

\bibitem[{\citenamefont{Ando}(2006)}]{AndoMac}
\bibinfo{author}{\bibfnamefont{T.}~\bibnamefont{Ando}}, \bibinfo{journal}{J.
  Phys. Soc. Jpn.} \textbf{\bibinfo{volume}{75}}, \bibinfo{pages}{074716}
  (\bibinfo{year}{2006}).

\bibitem[{\citenamefont{Nomura and MacDonald}(2007)}]{NomuraPRL}
\bibinfo{author}{\bibfnamefont{K.}~\bibnamefont{Nomura}} \bibnamefont{and}
  \bibinfo{author}{\bibfnamefont{A.~H.} \bibnamefont{MacDonald}},
  \bibinfo{journal}{Phys. Rev. Lett.} \textbf{\bibinfo{volume}{98}},
  \bibinfo{pages}{076602} (\bibinfo{year}{2007}).

\bibitem[{\citenamefont{Ponomarenko et~al.}(2009)\citenamefont{Ponomarenko,
  Yang, Mohiuddin, Katsnelson, Novoselov, Morozov, Zhukov, Schedin, Hill, and
  Geim}}]{GeimIce_PRL09}
\bibinfo{author}{\bibfnamefont{L.~A.} \bibnamefont{Ponomarenko}},
  \bibinfo{author}{\bibfnamefont{R.}~\bibnamefont{Yang}},
  \bibinfo{author}{\bibfnamefont{T.~M.} \bibnamefont{Mohiuddin}},
  \bibinfo{author}{\bibfnamefont{M.~I.} \bibnamefont{Katsnelson}},
  \bibinfo{author}{\bibfnamefont{K.~S.} \bibnamefont{Novoselov}},
  \bibinfo{author}{\bibfnamefont{S.~V.} \bibnamefont{Morozov}},
  \bibinfo{author}{\bibfnamefont{A.~A.} \bibnamefont{Zhukov}},
  \bibinfo{author}{\bibfnamefont{F.}~\bibnamefont{Schedin}},
  \bibinfo{author}{\bibfnamefont{E.~W.} \bibnamefont{Hill}}, \bibnamefont{and}
  \bibinfo{author}{\bibfnamefont{A.~K.} \bibnamefont{Geim}},
  \bibinfo{journal}{Phys. Rev. Lett.} \textbf{\bibinfo{volume}{102}},
  \bibinfo{pages}{206603} (\bibinfo{year}{2009}).

\bibitem[{\citenamefont{Schedin et~al.}(2007)\citenamefont{Schedin, Geim,
  Morozov, Hill, Blake, Katsnelson, and Novoselov}}]{Schedin_NM09}
\bibinfo{author}{\bibfnamefont{F.}~\bibnamefont{Schedin}},
  \bibinfo{author}{\bibfnamefont{A.~K.} \bibnamefont{Geim}},
  \bibinfo{author}{\bibfnamefont{S.~V.} \bibnamefont{Morozov}},
  \bibinfo{author}{\bibfnamefont{E.~W.} \bibnamefont{Hill}},
  \bibinfo{author}{\bibfnamefont{P.}~\bibnamefont{Blake}},
  \bibinfo{author}{\bibfnamefont{M.~I.} \bibnamefont{Katsnelson}},
  \bibnamefont{and} \bibinfo{author}{\bibfnamefont{K.~S.}
  \bibnamefont{Novoselov}}, \bibinfo{journal}{Nature Materials}
  \textbf{\bibinfo{volume}{6}}, \bibinfo{pages}{652} (\bibinfo{year}{2007}).

\bibitem[{\citenamefont{Li et~al.}(2011{\natexlab{b}})\citenamefont{Li, Hwang,
  Rossi, and Das~Sarma}}]{qzli_PRL11}
\bibinfo{author}{\bibfnamefont{Q.}~\bibnamefont{Li}},
  \bibinfo{author}{\bibfnamefont{E.~H.} \bibnamefont{Hwang}},
  \bibinfo{author}{\bibfnamefont{E.}~\bibnamefont{Rossi}}, \bibnamefont{and}
  \bibinfo{author}{\bibfnamefont{S.}~\bibnamefont{Das~Sarma}},
  \bibinfo{journal}{Phys. Rev. Lett.} \textbf{\bibinfo{volume}{107}},
  \bibinfo{pages}{156601} (\bibinfo{year}{2011}{\natexlab{b}}).

\bibitem[{\citenamefont{Rossi and Das~Sarma}(2008)}]{rossi2008}
\bibinfo{author}{\bibfnamefont{E.}~\bibnamefont{Rossi}} \bibnamefont{and}
  \bibinfo{author}{\bibfnamefont{S.}~\bibnamefont{Das~Sarma}},
  \bibinfo{journal}{Phys. Rev. Lett.} \textbf{\bibinfo{volume}{101}},
  \bibinfo{pages}{166803} (\bibinfo{year}{2008}).

\bibitem[{\citenamefont{Kawamura and Sarma}(1996)}]{Kawamura_SCC96}
\bibinfo{author}{\bibfnamefont{T.}~\bibnamefont{Kawamura}} \bibnamefont{and}
  \bibinfo{author}{\bibfnamefont{S.~D.} \bibnamefont{Sarma}},
  \bibinfo{journal}{Solid State Communications} \textbf{\bibinfo{volume}{100}},
  \bibinfo{pages}{411 } (\bibinfo{year}{1996}).

\bibitem[{\citenamefont{Caragiu and Finberg}(2005)}]{Caragiu_JPCM05}
\bibinfo{author}{\bibfnamefont{M.}~\bibnamefont{Caragiu}} \bibnamefont{and}
  \bibinfo{author}{\bibfnamefont{S.}~\bibnamefont{Finberg}},
  \bibinfo{journal}{J. Phys.: Condens. Matter} \textbf{\bibinfo{volume}{17}},
  \bibinfo{pages}{R995} (\bibinfo{year}{2005}).

\bibitem[{\citenamefont{Martin et~al.}(2008)\citenamefont{Martin, Akerman,
  Ulbricht, Lohmann, Smet, \mbox{von Klitzing}, and Yacobi}}]{martin2008}
\bibinfo{author}{\bibfnamefont{J.}~\bibnamefont{Martin}},
  \bibinfo{author}{\bibfnamefont{N.}~\bibnamefont{Akerman}},
  \bibinfo{author}{\bibfnamefont{G.}~\bibnamefont{Ulbricht}},
  \bibinfo{author}{\bibfnamefont{T.}~\bibnamefont{Lohmann}},
  \bibinfo{author}{\bibfnamefont{J.~H.} \bibnamefont{Smet}},
  \bibinfo{author}{\bibfnamefont{K.}~\bibnamefont{\mbox{von Klitzing}}},
  \bibnamefont{and} \bibinfo{author}{\bibfnamefont{A.}~\bibnamefont{Yacobi}},
  \bibinfo{journal}{Nature Physics} \textbf{\bibinfo{volume}{4}},
  \bibinfo{pages}{144} (\bibinfo{year}{2008}).

\bibitem[{\citenamefont{Zhang et~al.}(2009)\citenamefont{Zhang, Brar, Girit,
  Zettl, and Crommie}}]{YZhang_NP09}
\bibinfo{author}{\bibfnamefont{Y.}~\bibnamefont{Zhang}},
  \bibinfo{author}{\bibfnamefont{V.~W.} \bibnamefont{Brar}},
  \bibinfo{author}{\bibfnamefont{C.}~\bibnamefont{Girit}},
  \bibinfo{author}{\bibfnamefont{A.}~\bibnamefont{Zettl}}, \bibnamefont{and}
  \bibinfo{author}{\bibfnamefont{M.~F.} \bibnamefont{Crommie}},
  \bibinfo{journal}{Nat. Phys.} \textbf{\bibinfo{volume}{5}},
  \bibinfo{pages}{722} (\bibinfo{year}{2009}).

\bibitem[{\citenamefont{Deshpande
  et~al.}(2009{\natexlab{a}})\citenamefont{Deshpande, Bao, Miao, Lau, and
  LeRoy}}]{LeRoyPuddle_PRB09}
\bibinfo{author}{\bibfnamefont{A.}~\bibnamefont{Deshpande}},
  \bibinfo{author}{\bibfnamefont{W.}~\bibnamefont{Bao}},
  \bibinfo{author}{\bibfnamefont{F.}~\bibnamefont{Miao}},
  \bibinfo{author}{\bibfnamefont{C.~N.} \bibnamefont{Lau}}, \bibnamefont{and}
  \bibinfo{author}{\bibfnamefont{B.~J.} \bibnamefont{LeRoy}},
  \bibinfo{journal}{Phys. Rev. B} \textbf{\bibinfo{volume}{79}},
  \bibinfo{pages}{205411} (\bibinfo{year}{2009}{\natexlab{a}}).

\bibitem[{\citenamefont{Adam et~al.}(2009)\citenamefont{Adam, Hwang, Rossi, and
  Sarma}}]{Adam_SCC09}
\bibinfo{author}{\bibfnamefont{S.}~\bibnamefont{Adam}},
  \bibinfo{author}{\bibfnamefont{E.}~\bibnamefont{Hwang}},
  \bibinfo{author}{\bibfnamefont{E.}~\bibnamefont{Rossi}}, \bibnamefont{and}
  \bibinfo{author}{\bibfnamefont{S.~D.} \bibnamefont{Sarma}},
  \bibinfo{journal}{Solid State Communications} \textbf{\bibinfo{volume}{149}},
  \bibinfo{pages}{1072 } (\bibinfo{year}{2009}).

\bibitem[{\citenamefont{Deshpande et~al.}(2011)\citenamefont{Deshpande, Bao,
  Zhao, Lau, and LeRoy}}]{LeRoyPuddle_PRB11}
\bibinfo{author}{\bibfnamefont{A.}~\bibnamefont{Deshpande}},
  \bibinfo{author}{\bibfnamefont{W.}~\bibnamefont{Bao}},
  \bibinfo{author}{\bibfnamefont{Z.}~\bibnamefont{Zhao}},
  \bibinfo{author}{\bibfnamefont{C.~N.} \bibnamefont{Lau}}, \bibnamefont{and}
  \bibinfo{author}{\bibfnamefont{B.~J.} \bibnamefont{LeRoy}},
  \bibinfo{journal}{Phys. Rev. B} \textbf{\bibinfo{volume}{83}},
  \bibinfo{pages}{155409} (\bibinfo{year}{2011}).

\bibitem[{\citenamefont{Hwang and Das~Sarma}(2007)}]{HwangDas_PRB07}
\bibinfo{author}{\bibfnamefont{E.~H.} \bibnamefont{Hwang}} \bibnamefont{and}
  \bibinfo{author}{\bibfnamefont{S.}~\bibnamefont{Das~Sarma}},
  \bibinfo{journal}{Phys. Rev. B} \textbf{\bibinfo{volume}{75}},
  \bibinfo{pages}{205418} (\bibinfo{year}{2007}).

\bibitem[{not()}]{note01}
\bibinfo{note}{For densities below $n=5\times 10^{12}{\rm cm}^{-2}$ the value
  of the conductivity obtained using the Boltzmann theory depends very weakly
  on $d$ ($\sigma$ changes by less than 10\% in going from $d=0$ to $d=1$ nm)
  and therefore in the remainder we set d=0 to simplify the analytical
  expressions for the relaxation time and $\sigma$, see also
  Ref.~\onlinecite{dassarma2010}.}

\bibitem[{\citenamefont{Hwang and
  Das~Sarma}(2008{\natexlab{c}})}]{Hwang_PRB195412}
\bibinfo{author}{\bibfnamefont{E.~H.} \bibnamefont{Hwang}} \bibnamefont{and}
  \bibinfo{author}{\bibfnamefont{S.}~\bibnamefont{Das~Sarma}},
  \bibinfo{journal}{Phys. Rev. B} \textbf{\bibinfo{volume}{77}},
  \bibinfo{pages}{195412} (\bibinfo{year}{2008}{\natexlab{c}}).

\bibitem[{\citenamefont{Yan and Fuhrer}(2011)}]{fuhrer2010}
\bibinfo{author}{\bibfnamefont{J.}~\bibnamefont{Yan}} \bibnamefont{and}
  \bibinfo{author}{\bibfnamefont{M.~S.} \bibnamefont{Fuhrer}},
  \bibinfo{journal}{Phys. Rev. Lett.} \textbf{\bibinfo{volume}{107}},
  \bibinfo{pages}{206601} (\bibinfo{year}{2011}).

\bibitem[{\citenamefont{Bruggeman}(1935)}]{bruggeman1935}
\bibinfo{author}{\bibfnamefont{D.~A.~G.} \bibnamefont{Bruggeman}},
  \bibinfo{journal}{Ann. Physik} \textbf{\bibinfo{volume}{416}},
  \bibinfo{pages}{636} (\bibinfo{year}{1935}).

\bibitem[{\citenamefont{Landauer}(1952)}]{landauer1952}
\bibinfo{author}{\bibfnamefont{R.}~\bibnamefont{Landauer}},
  \bibinfo{journal}{J. Appl. Phys.} \textbf{\bibinfo{volume}{23}},
  \bibinfo{pages}{779} (\bibinfo{year}{1952}).

\bibitem[{\citenamefont{Landauer}(1978)}]{landauer1978}
\bibinfo{author}{\bibfnamefont{R.}~\bibnamefont{Landauer}}, in
  \emph{\bibinfo{booktitle}{Electrical transport and optical properties of
  inhomogeneous media.}}, edited by \bibinfo{editor}{\bibfnamefont{J.~C.}
  \bibnamefont{Garland}} \bibnamefont{and}
  \bibinfo{editor}{\bibfnamefont{D.~B.} \bibnamefont{Tanner}}
  (\bibinfo{year}{1978}), p.~\bibinfo{pages}{2}.

\bibitem[{\citenamefont{Fogler}(2009)}]{fogler2009}
\bibinfo{author}{\bibfnamefont{M.~M.} \bibnamefont{Fogler}},
  \bibinfo{journal}{Phys. Rev. Lett.} \textbf{\bibinfo{volume}{103}},
  \bibinfo{pages}{236801} (\bibinfo{year}{2009}).

\bibitem[{\citenamefont{Sarma et~al.}()\citenamefont{Sarma, Hwang, and
  Li}}]{DasHwangLi_arXiv11}
\bibinfo{author}{\bibfnamefont{S.~D.} \bibnamefont{Sarma}},
  \bibinfo{author}{\bibfnamefont{E.~H.} \bibnamefont{Hwang}}, \bibnamefont{and}
  \bibinfo{author}{\bibfnamefont{Q.}~\bibnamefont{Li}},
  \bibinfo{note}{arXiv:1109.0988 (2011)}.

\bibitem[{\citenamefont{McCann et~al.}(2006)\citenamefont{McCann, Kechedzhi,
  Fal'ko, Suzuura, Ando, and Altshuler}}]{mccann2006}
\bibinfo{author}{\bibfnamefont{E.}~\bibnamefont{McCann}},
  \bibinfo{author}{\bibfnamefont{K.}~\bibnamefont{Kechedzhi}},
  \bibinfo{author}{\bibfnamefont{V.~I.} \bibnamefont{Fal'ko}},
  \bibinfo{author}{\bibfnamefont{H.}~\bibnamefont{Suzuura}},
  \bibinfo{author}{\bibfnamefont{T.}~\bibnamefont{Ando}}, \bibnamefont{and}
  \bibinfo{author}{\bibfnamefont{B.}~\bibnamefont{Altshuler}},
  \bibinfo{journal}{Phys. Rev. Lett.} \textbf{\bibinfo{volume}{97}},
  \bibinfo{pages}{146805} (\bibinfo{year}{2006}).

\bibitem[{\citenamefont{Castro et~al.}(2007)\citenamefont{Castro, Novoselov,
  Morozov, Peres, dos Santos, Nilsson, Guinea, Geim, and Neto}}]{castro2007}
\bibinfo{author}{\bibfnamefont{E.~V.} \bibnamefont{Castro}},
  \bibinfo{author}{\bibfnamefont{K.~S.} \bibnamefont{Novoselov}},
  \bibinfo{author}{\bibfnamefont{S.~V.} \bibnamefont{Morozov}},
  \bibinfo{author}{\bibfnamefont{N.~M.~R.} \bibnamefont{Peres}},
  \bibinfo{author}{\bibfnamefont{J.~M. B.~L.} \bibnamefont{dos Santos}},
  \bibinfo{author}{\bibfnamefont{J.}~\bibnamefont{Nilsson}},
  \bibinfo{author}{\bibfnamefont{F.}~\bibnamefont{Guinea}},
  \bibinfo{author}{\bibfnamefont{A.~K.} \bibnamefont{Geim}}, \bibnamefont{and}
  \bibinfo{author}{\bibfnamefont{A.~H.~C.} \bibnamefont{Neto}},
  \bibinfo{journal}{Phys. Rev. Lett.} \textbf{\bibinfo{volume}{99}},
  \bibinfo{pages}{216802} (\bibinfo{year}{2007}).

\bibitem[{\citenamefont{Oostinga et~al.}(2008)\citenamefont{Oostinga, Heersche,
  Liu, Morpurgo, and Vandersypen}}]{oostinga2008}
\bibinfo{author}{\bibfnamefont{J.~B.} \bibnamefont{Oostinga}},
  \bibinfo{author}{\bibfnamefont{H.~B.} \bibnamefont{Heersche}},
  \bibinfo{author}{\bibfnamefont{X.}~\bibnamefont{Liu}},
  \bibinfo{author}{\bibfnamefont{A.~F.} \bibnamefont{Morpurgo}},
  \bibnamefont{and} \bibinfo{author}{\bibfnamefont{L.~M.~K.}
  \bibnamefont{Vandersypen}}, \bibinfo{journal}{Nat. Mater.}
  \textbf{\bibinfo{volume}{7}}, \bibinfo{pages}{151} (\bibinfo{year}{2008}).

\bibitem[{\citenamefont{Mak et~al.}(2009)\citenamefont{Mak, Lui, Shan, and
  Heinz}}]{mak2009}
\bibinfo{author}{\bibfnamefont{K.~F.} \bibnamefont{Mak}},
  \bibinfo{author}{\bibfnamefont{C.~H.} \bibnamefont{Lui}},
  \bibinfo{author}{\bibfnamefont{J.}~\bibnamefont{Shan}}, \bibnamefont{and}
  \bibinfo{author}{\bibfnamefont{T.~F.} \bibnamefont{Heinz}},
  \bibinfo{journal}{Phys. Rev. Lett.} \textbf{\bibinfo{volume}{102}},
  \bibinfo{pages}{256405} (\bibinfo{year}{2009}).

\bibitem[{\citenamefont{Zou and Zhu}(2010)}]{zhujun_PRBR10}
\bibinfo{author}{\bibfnamefont{K.}~\bibnamefont{Zou}} \bibnamefont{and}
  \bibinfo{author}{\bibfnamefont{J.}~\bibnamefont{Zhu}},
  \bibinfo{journal}{Phys. Rev. B} \textbf{\bibinfo{volume}{82}},
  \bibinfo{pages}{081407} (\bibinfo{year}{2010}).

\bibitem[{\citenamefont{Rossi and Das~Sarma}(2011)}]{Rossi_PRL11}
\bibinfo{author}{\bibfnamefont{E.}~\bibnamefont{Rossi}} \bibnamefont{and}
  \bibinfo{author}{\bibfnamefont{S.}~\bibnamefont{Das~Sarma}},
  \bibinfo{journal}{Phys. Rev. Lett.} \textbf{\bibinfo{volume}{107}},
  \bibinfo{pages}{155502} (\bibinfo{year}{2011}).

\bibitem[{\citenamefont{Hwang and
  Das~Sarma}(2008{\natexlab{d}})}]{HwangBG_PRL08}
\bibinfo{author}{\bibfnamefont{E.~H.} \bibnamefont{Hwang}} \bibnamefont{and}
  \bibinfo{author}{\bibfnamefont{S.}~\bibnamefont{Das~Sarma}},
  \bibinfo{journal}{Phys. Rev. Lett.} \textbf{\bibinfo{volume}{101}},
  \bibinfo{pages}{156802} (\bibinfo{year}{2008}{\natexlab{d}}).

\bibitem[{\citenamefont{Das~Sarma et~al.}(2010)\citenamefont{Das~Sarma, Hwang,
  and Rossi}}]{DasEnrico_PRB10}
\bibinfo{author}{\bibfnamefont{S.}~\bibnamefont{Das~Sarma}},
  \bibinfo{author}{\bibfnamefont{E.~H.} \bibnamefont{Hwang}}, \bibnamefont{and}
  \bibinfo{author}{\bibfnamefont{E.}~\bibnamefont{Rossi}},
  \bibinfo{journal}{Phys. Rev. B} \textbf{\bibinfo{volume}{81}},
  \bibinfo{pages}{161407} (\bibinfo{year}{2010}).

\bibitem[{\citenamefont{Adam and {Das Sarma}}(2008)}]{adam2008a}
\bibinfo{author}{\bibfnamefont{S.}~\bibnamefont{Adam}} \bibnamefont{and}
  \bibinfo{author}{\bibfnamefont{S.}~\bibnamefont{{Das Sarma}}},
  \bibinfo{journal}{Phys. Rev. B} \textbf{\bibinfo{volume}{77}},
  \bibinfo{pages}{115436} (\bibinfo{year}{2008}).

\bibitem[{\citenamefont{Deshpande
  et~al.}(2009{\natexlab{b}})\citenamefont{Deshpande, Bao, Zhao, Lau, and
  LeRoy}}]{deshpande2009b}
\bibinfo{author}{\bibfnamefont{A.}~\bibnamefont{Deshpande}},
  \bibinfo{author}{\bibfnamefont{W.}~\bibnamefont{Bao}},
  \bibinfo{author}{\bibfnamefont{Z.}~\bibnamefont{Zhao}},
  \bibinfo{author}{\bibfnamefont{C.~N.} \bibnamefont{Lau}}, \bibnamefont{and}
  \bibinfo{author}{\bibfnamefont{B.~J.} \bibnamefont{LeRoy}},
  \bibinfo{journal}{Appl. Phys. Lett.} \textbf{\bibinfo{volume}{95}},
  \bibinfo{pages}{243502} (\bibinfo{year}{2009}{\natexlab{b}}).

\bibitem[{\citenamefont{Das~Sarma and Hwang}(2011)}]{DasHwang_BN11}
\bibinfo{author}{\bibfnamefont{S.}~\bibnamefont{Das~Sarma}} \bibnamefont{and}
  \bibinfo{author}{\bibfnamefont{E.~H.} \bibnamefont{Hwang}},
  \bibinfo{journal}{Phys. Rev. B} \textbf{\bibinfo{volume}{83}},
  \bibinfo{pages}{121405} (\bibinfo{year}{2011}).

\end{thebibliography}
\end{document}